\documentclass[prl, twocolumn, superscriptaddress, 10pt]{revtex4}  
\usepackage{amssymb}
\usepackage{amsmath}
\usepackage{bbm}
\usepackage{graphicx}

\usepackage{color}

\newcommand{\hc}{\hat{c}}

\newcommand{\hd}{\hat{d}}
\newcommand{\hH}{\hat{H}}
\newcommand{\hh}{\hat{h}}
\newcommand{\hn}{\hat{n}}

\newcommand{\hD}{\hat{D}}
\newcommand{\hS}{\hat{S}}
\newcommand{\hrho}{\hat{\rho}}
\newcommand{\brho}{\boldsymbol{\rho}}
\newcommand{\hbrho}{\hat{\boldsymbol {\rho}}}
\newcommand{\htau}{\hat{\tau}}
\newcommand{\btau}{\boldsymbol{\tau}}
\newcommand{\hbtau}{\hat{\boldsymbol {\tau}}}

\newcommand{\eqq}[1]{\begin{align} #1 \end{align}}

\begin{document}

\author{Philipp Werner}
\affiliation{Department of Physics, University of Fribourg, 1700 Fribourg, Switzerland}
\author{Yuta Murakami}
\affiliation{Department of Physics, Tokyo Institute of Technology, Meguro, Tokyo 152-8551, Japan}

\title{Nonthermal excitonic condensation near a spin-state transition}

\date{\today}

\hyphenation{}

\begin{abstract}
We consider a two-orbital Hubbard model with Hund coupling and crystal-field splitting and show that in the vicinity of the high-spin/low-spin transition, crystal-field quenches can induce an excitonic condensation at initial temperatures above the highest ordering temperature in equilibrium. This condensation is the effect of an increase in the spin entropy and an associated cooling of the effective electronic temperature. We identify a dynamical phase transition and show that such quenches can result in long-lived nonthermal excitonic condensates which have no analogue in the equilibrium phase diagram. The results are interpreted by means of an effective pseudo-spin model. 
\end{abstract}

%\pacs{ 71.10.Fd}

\maketitle

The nonequilibrium control of electronic orders in correlated lattice systems and materials is an intriguing prospect with potentially important technological applications. Several recent experiments and numerical simulations suggest that photo-excitation can induce or enhance superconducting \cite{Fausti_2011,Kaiser_2014,Denny_2015,Mitrano_2016,Sentef2016,Okamoto_2016,Kennes2017,Mazza_2017,Werner2018,Nava_2018,Kaneko_2019,Werner_2019b,Kennes_2019,Tindall_2019,Buzzi_2020},  
excitonic \cite{Mor_2017,Murakami_2017,Ohta_2018,Yonemitsu_2018,Murotani_2019,Stefanucci_2019,Murakami_2020}, or magnetic order \cite{Ono_2017,Werner_2019a}. One possible strategy, which has been theoretically explored in different contexts \cite{Bernier_2009,Denny_2015,Nava_2018,Fabrizio2018,Werner_2019a,Werner_2019b}, and successfully employed in cold atom experiments \cite{Mazurenko_2017,Chiu_2018}, is to reshuffle entropy between different subsystems to cool down the relevant degrees of freedom. In particular, the separation into high-density and low-density regions \cite{Bernier_2009} and charge excitations between different bands \cite{Werner_2019a,Werner_2019b} 
result in a substantial entropy transfer. The excitation of spin-triplet excitons has also been suggested as a possible entropy sink in photo-excited fulleride superconductors \cite{Nava_2018}. 

An unexplored playground for studying entropic effects on nonequilibrium dynamics are multi-orbital Hubbard systems near spin-state transitions, where complex ordering phenomena, including excitonic insulator (EI) phases, exist in equilibrium \cite{Kaneko_2012,Kunes_2014,Kunes_2015,Nasu_2016,Hoshino_2016}.
A transition, e.~g. from a paramagnetic spin-0 insulator to a paramagnetic spin-1 insulator, is associated with a dramatic change in the spin entropy (an increase by $\sim \ln(3)$ per site). If such a transition were accomplished without increase in the total entropy, it would trigger a substantial cooling of the electrons and, possibly, electronic ordering instabilities. 
In this letter, we show that this entropy cooling mechanism is very effective, even in the case of quenches, and that it allows to realize an EI in systems with initial temperatures above the highest equilibrium ordering temperature. Even more interestingly, we show that quenches in the vicinity of spin-state transitions allow to induce a nonthermal form of excitonic order which has no correspondence in the equilibrium phase diagram. 

We consider a two-orbital Hubbard model with local interaction 
$H_\text{int}=\sum_{a} U n_{a,\uparrow}n_{a,\downarrow}+\sum_{\sigma} [U' n_{1,\sigma}n_{2,\bar\sigma}+(U'-J) n_{1,\sigma}n_{2,\sigma}]-J(c^\dagger_{1\downarrow}c^\dagger_{2\uparrow}c_{2\downarrow}c_{1\uparrow} + \text{h.c.})+I(c^\dagger_{2\uparrow}c^\dagger_{2\downarrow}c_{1\uparrow}c_{1\downarrow} + \text{h.c.})$ 
parametrized by the intra-orbital interaction $U$, Hund coupling $J$ ($U'=U-2J$), and pair-hopping interaction  $I$, and a crystal-field splitting $H_\text{cf}=\Delta_\text{cf} (n_1-n_2),\label{cf}$
where $a=1,2$ denotes the orbitals, $\sigma$ the spin, $n_{a\sigma}=c^\dagger_{a\sigma}c_{a\sigma}$ the spin and orbital dependent density, and $n_a=(n_{a\uparrow}+n_{a\downarrow})$. There is an orbital-diagonal nearest-neighbor hopping $v$ and an orbital-offdiagonal hopping $v'$. 
We treat the lattice model within the dynamical mean field theory (DMFT) approximation \cite{Georges_1996} assuming an infinitely connected Bethe lattice. 
In this case the hybridization function matrix ${\bf \Delta}(t,t')$ of the DMFT impurity problem is determined directly by the local Green's function matrix ${\bf G}(t,t')$ through ${\bf \Delta}(t,t') = {\bf V} {\bf G}(t,t') {\bf V}$, where 
${\bf G}(t,t')=-i\langle T_\mathcal{C} \psi(t)\psi^\dagger(t')\rangle$ \cite{Aoki_2014} with $\psi^\dagger=(c^\dagger_{1\uparrow}, c^\dagger_{2\uparrow}, c^\dagger_{1\downarrow}, c^\dagger_{2\downarrow})$, and
${\bf V} = \text{diag}({\bf v}, {\bf v})$. 
Here, ${\bf v}$ denotes a $2\times 2$ matrix with $(v,-v)$ on the diagonal and $v'$ on the offdiagonal. 
In the following we set $I=J$ and $v'=v$, and use $v=1$ as the unit of energy.  
The DMFT impurity problem is solved by the non-crossing approximation (NCA) \cite{Keiter1971,Eckstein2010}.

\begin{figure*}[t]
\begin{center}
\includegraphics[angle=-90, width=0.329\textwidth]{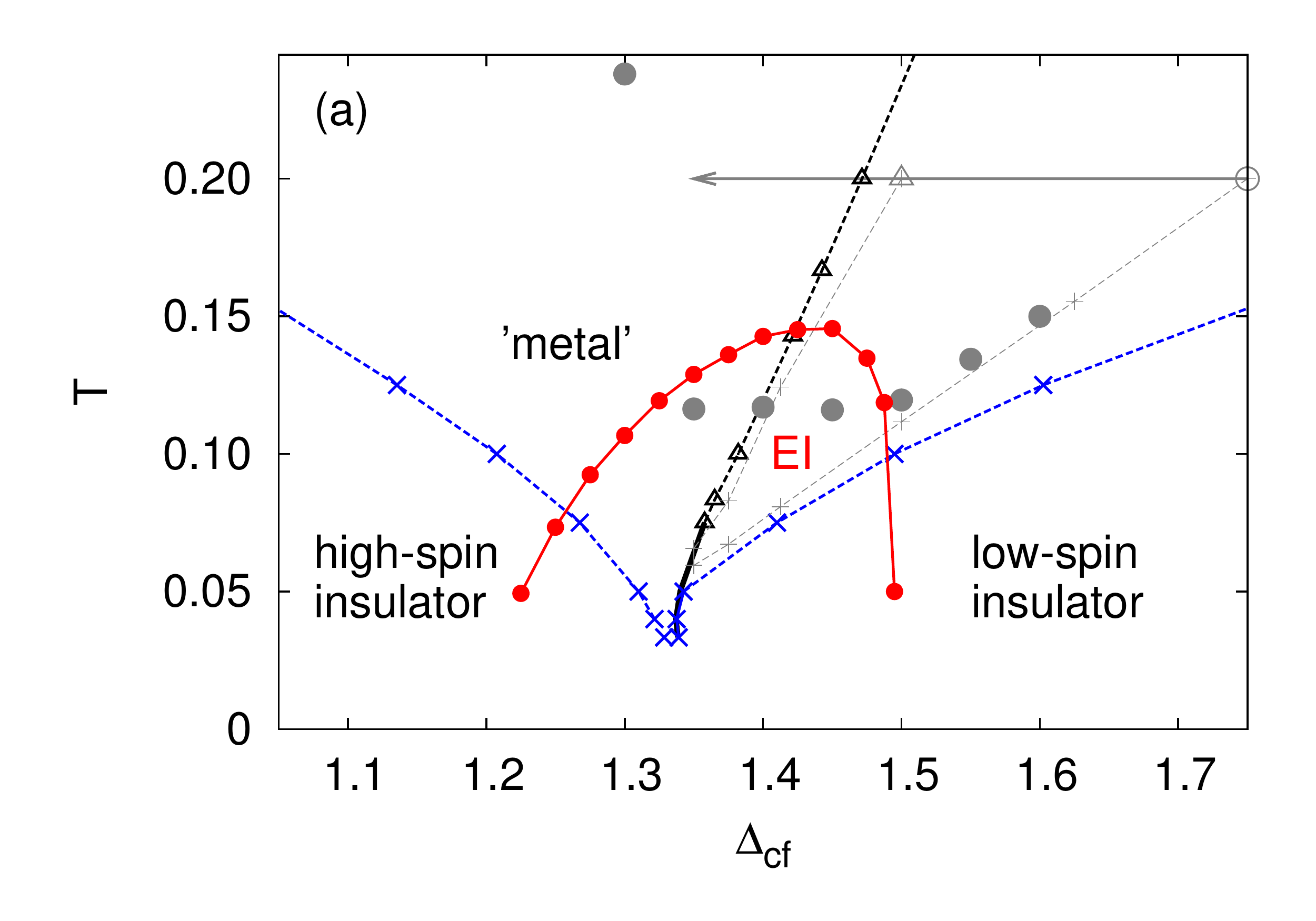}\hfill
\includegraphics[angle=-90, width=0.329\textwidth]{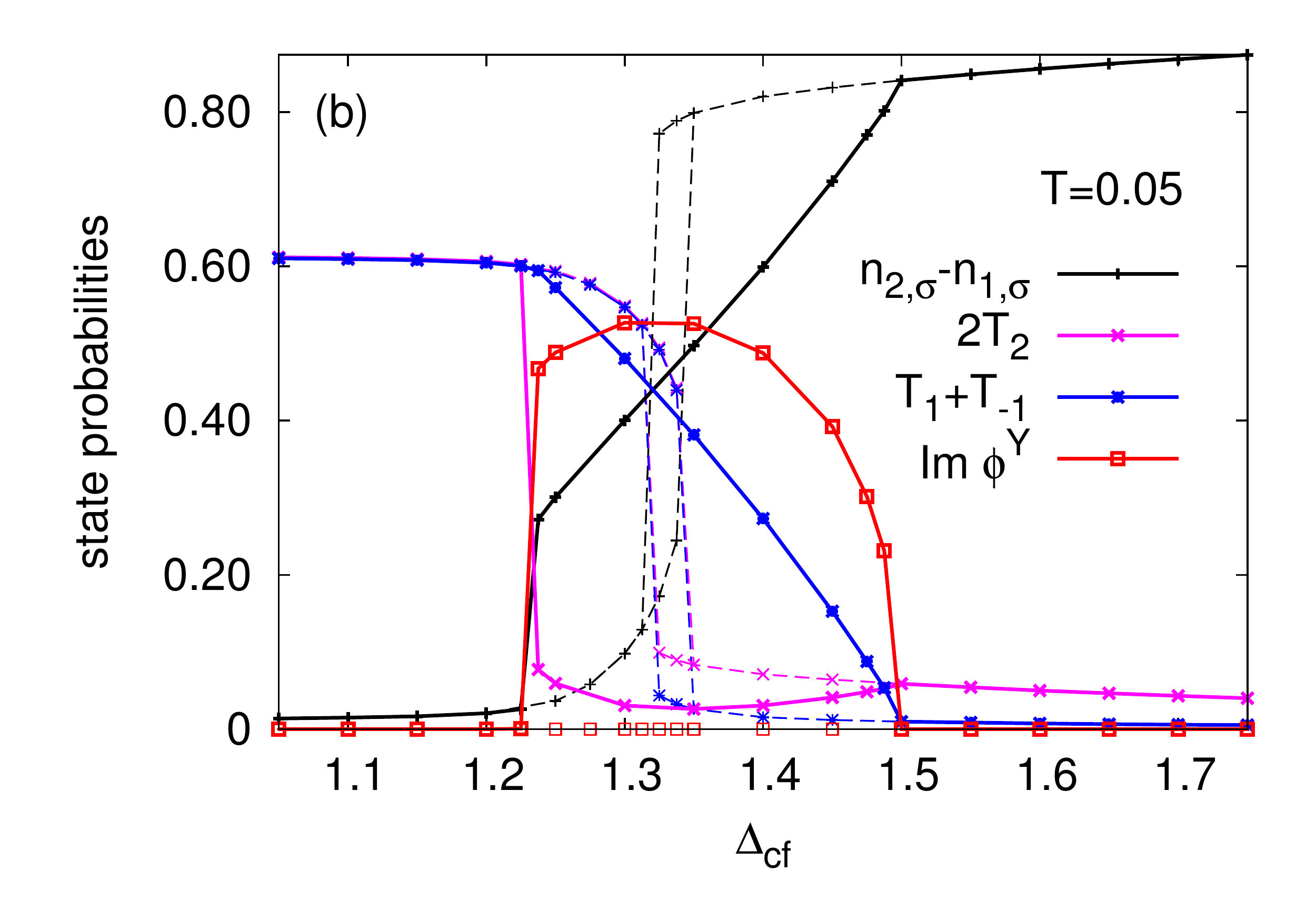}\hfill
\includegraphics[angle=-90, width=0.329\textwidth]{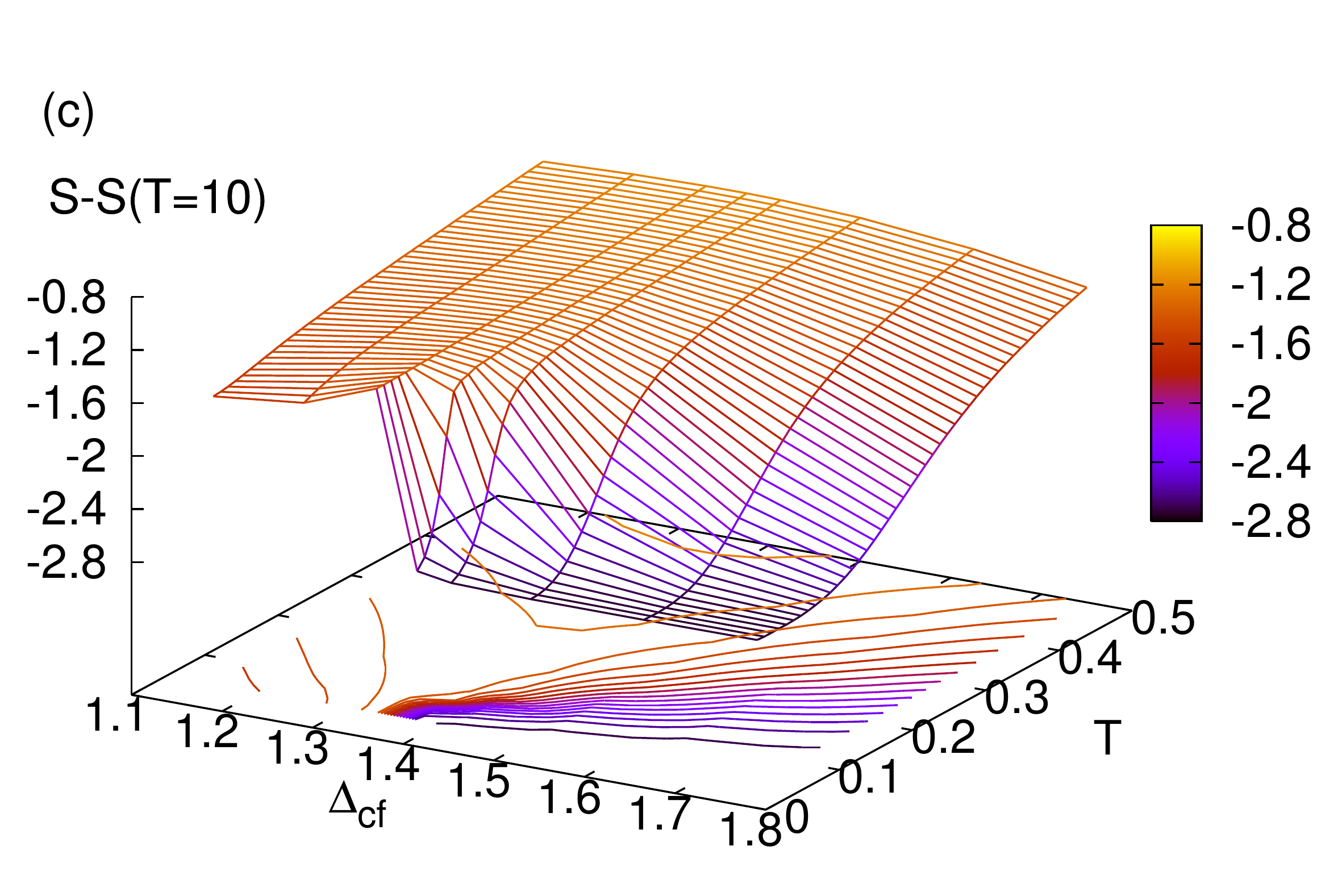} % 60 90 514 770
\caption{(a) Phase diagram of the half-filled model for $U=6$ and $J=1$. The black lines show the high-spin/low-spin transition (solid) or crossover (dashed), while blue dashed lines indicate the insulator-metal crossover defined by $\tfrac1T G_{a\sigma}(\tfrac{1}{2T})=0.033$. Red dots indicate the EI phase. (b) Orbital polarization (black), relevant local states (pink and blue, see text for the definition), and EI OP (red) as a function of crystal-field splitting at $T=0.05$. The first order transition with hysteresis in the absence of symmetry breaking is shown by the dashed lines.
(c) Entropy per site in the symmetric phase, relative to the value at $T=10$. The bottom plane shows the constant entropy contours.
}
\label{fig_phasediagram}
\end{center}
\end{figure*}

We first discuss the equilibrium phase diagram of this two-orbital model at half-filling. The competition between the Hund coupling and crystal-field splitting leads, in the strongly correlated regime and at low temperatures, to a transition between a high-spin Mott insulator with approximately one electron per orbital and a low-spin insulator (LI) with almost two electrons in the lower orbital \cite{Werner_2007}. Kunes and co-workers showed that the region in the vicinity of the spin-state transition (or crossover) is unstable to staggered high-spin/low-spin or excitonic order \cite{Kunes_2014,Kunes_2015}. Following Ref.~\onlinecite{Kunes_2015} we define the spin-triplet excitonic order parameters (OPs) as 
\begin{equation}
\phi^\lambda = \sum_{\sigma\sigma'} \langle c^\dagger_{1\sigma} c_{2\sigma'} \rangle {\bf \sigma}_{\sigma\sigma'}^\lambda,
\end{equation}
where ${\bf \sigma}_{\sigma\sigma'}^\lambda$ denotes the Pauli matrix for $\lambda=X,Y,Z$ \cite{footnote_op}.

The phase diagram of our model in the space of crystal-field splitting $\Delta_\text{cf}$ and temperature $T$ is shown in Fig.~\ref{fig_phasediagram}(a) for $U=6$ and $J=1$. Let us first discuss the transitions and crossovers in the absence of spontaneous symmetry breaking (blue and black lines). In the atomic limit, the level crossing associated with the high-spin/low-spin transition occurs at $\Delta_\text{cf}=\sqrt{2}J$ \cite{Werner_2007}. In the DMFT+NCA solution, it is shifted to slightly lower values ($\approx 1.34$ at low $T$). The solid black line corresponds to a first order phase transition, while the dashed line at $T\gtrsim 0.075$ indicates a high-spin/low-spin crossover. 

If we allow symmetry breaking to spin-triplet excitonic order, the low-temperature region around the high-spin/low-spin transition exhibits a spontaneous formation of an inter-orbital hybridization \cite{Kunes_2014,Kunes_2015}. $\phi^X$,$\phi^Y$ and $\phi^Z$ are equivalent by symmetry, and we consider  excitonic order with nonzero $\phi^Y$ and zero $\phi^{X,Z}$, without loss of generality. Since we apply a small seed field in the imaginary $\phi^Y$ direction, the 
EI phase (indicated by the red line in the phase diagram) corresponds to a purely imaginary $\phi^Y$ (or real $\langle c^\dagger_{1\uparrow}c_{2\downarrow}-c^\dagger_{1\downarrow}c_{2\uparrow}\rangle$). 
In Fig.~\ref{fig_phasediagram}(b), we plot the weight of the relevant local states and the OP for fixed $T=0.05$ as a function of $\Delta_\text{cf}$. Dashed lines show the results in the absence of symmetry breaking,  with a small hysteresis region around the first order transition, and full lines those in the presence of excitonic order. The black line indicates the orbital polarization, $n_{1,\sigma}-n_{2,\sigma}$, the pink line marked with ``$2T_2$" the combined weight of the $|\!\!\uparrow,\downarrow\rangle$ ($=\hat{c}^\dagger_{1\uparrow}\hat{c}^\dagger_{2\downarrow}|{\rm vac\rangle}$) and $|\!\!\downarrow,\uparrow\rangle$ states, and the blue line marked with ``$T_1+T_{-1}$" the combined weight of the $|\!\uparrow,\uparrow\rangle$ and $|\!\downarrow,\downarrow\rangle$ states. 
The high-spin insulator (HI) is essentially dominated by the three triplet states, while the orbital polarization 
in the LI  is very large. The transition from the LI to the EI seems to be continuous, 
while there is a large jump in the OP and the $T_2$ weight at the boundary between the EI and HI (for the evolution of the spectral functions across these transitions, see Supplementary Material (SM)). 
Note that $T_{1,-1}\ll T_2$ since we consider the EI with dominant $\phi^Y$ \cite{footnote_phiz}.

\begin{figure*}[t]
\begin{center}
\includegraphics[angle=-90, width=0.325\textwidth]{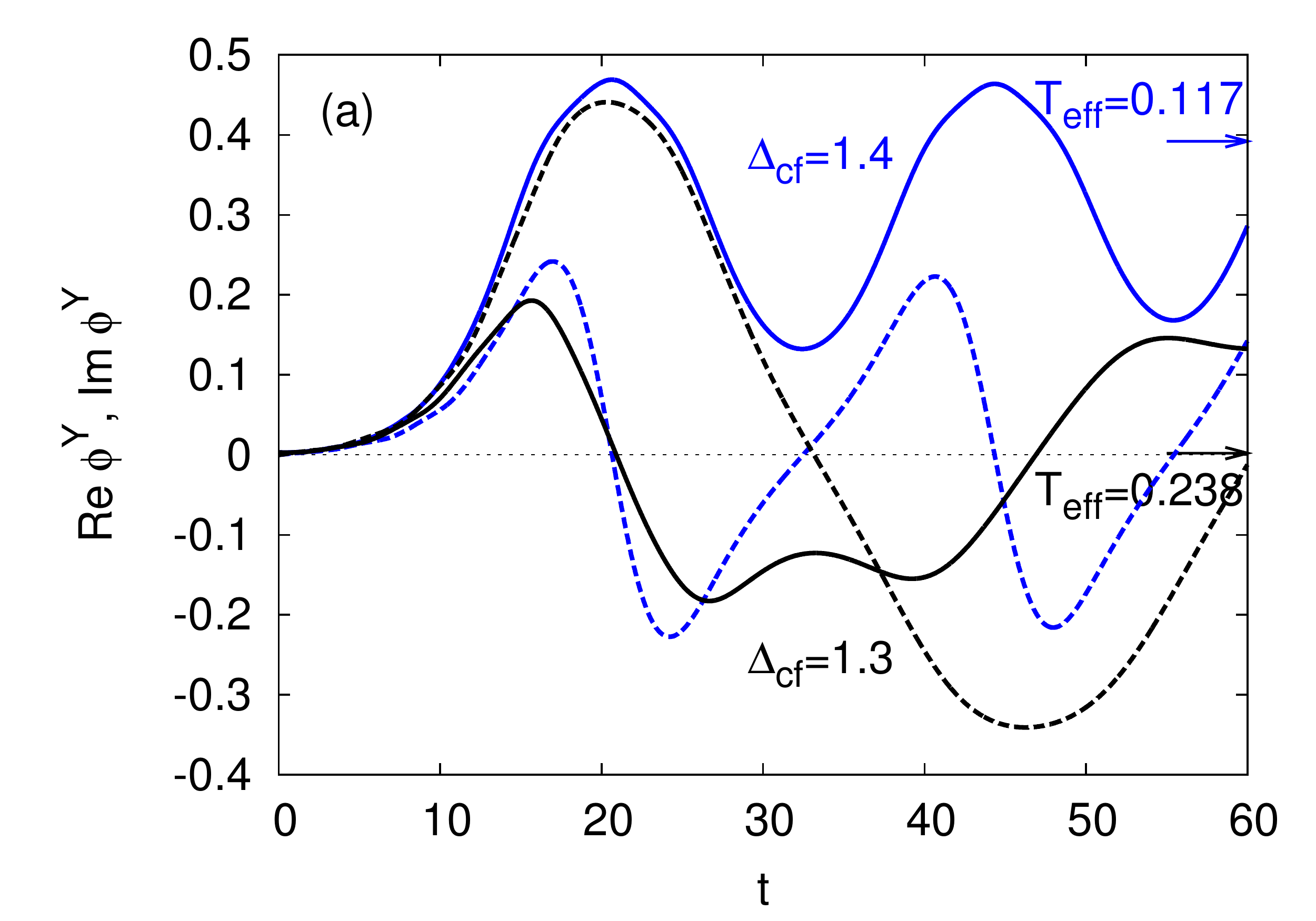}%\hfill
\includegraphics[angle=-90, width=0.345\textwidth]{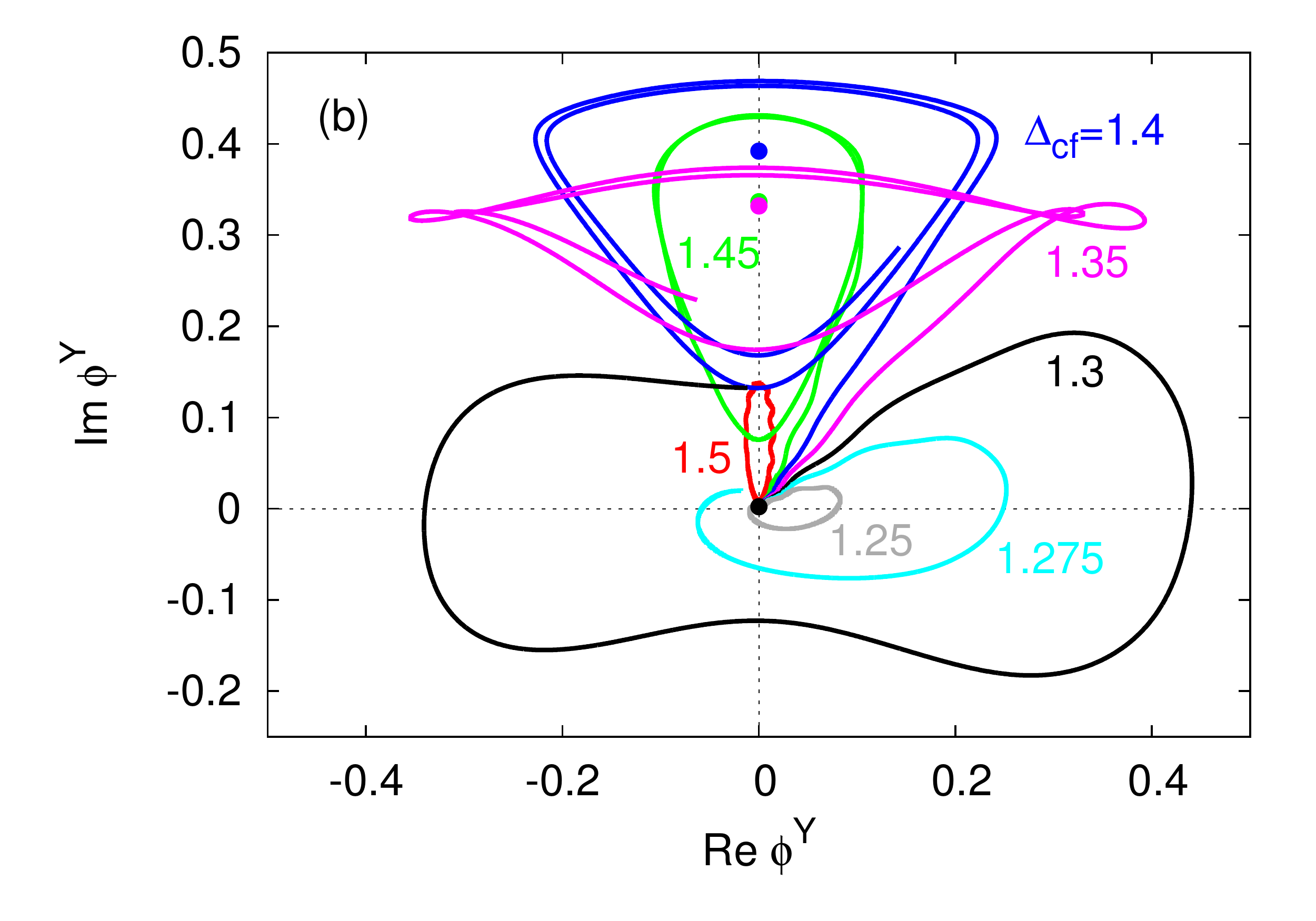}\hfill
\includegraphics[angle=-90, width=0.325\textwidth]{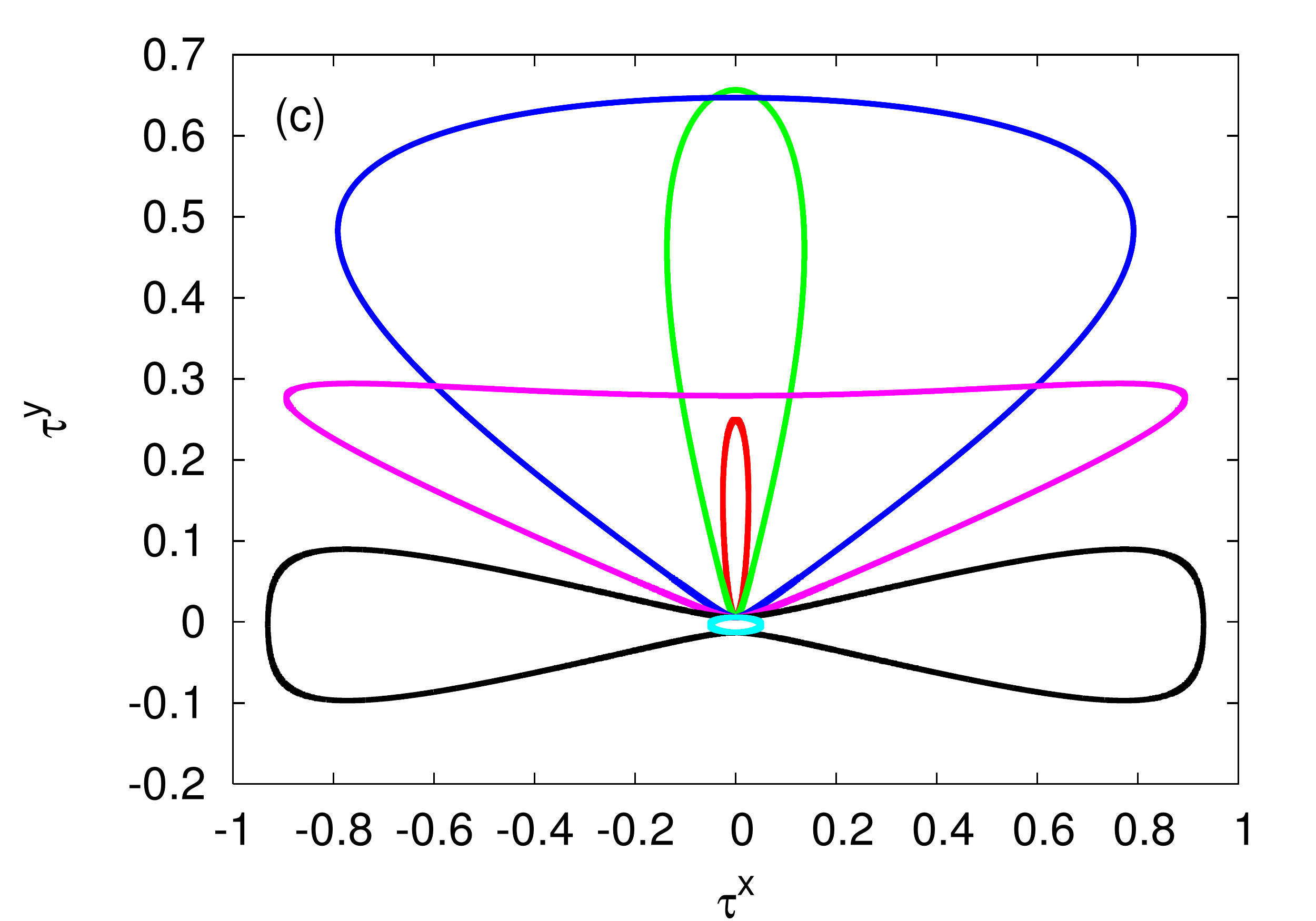}
\caption{(a,b) Evolution of the OP after a quench from $\Delta_\text{cf}=1.75$, $T=0.2$ to the indicated values of $\Delta_\text{cf}$. In panel (a) we plot the real (dashed lines) and imaginary (solid lines) part of $\phi^Y(t)$, and in (b) the trajectory of the OP in the complex plane. Arrows and colored dots indicate the values reached after thermalization. Panel (c) illustrates the MF dynamics for a quench from $\Delta_\text{cf}=2.0$, $T=0.3$ to $\Delta_\text{cf}=1.74$, 1.68, 1.39, 1.345, 1.325. $\tau^x_Y\approx \sqrt{2}\text{Re}\phi^Y$ and $\tau^y_Y\approx \sqrt{2}\text{Im}\phi^Y$.}
\label{fig_phiy}
\end{center}
\end{figure*}

To interpret the nonequilibrium results it is important to discuss the entropy of the system. Since the LI is spin-singlet and the HI spin-triplet, we expect a substantial change in the spin entropy near the high-spin/low-spin transition or crossover. The entropy can be obtained by integrating $C_V/T$ (with $C_V=dE_\text{tot}/dT$ the specific heat) from high temperatures: $S(T)=S_\infty-\int_T^\infty C_V(T')/T' dT'$. In DMFT, the total energy per site can be calculated as $E_\text{tot}=\langle H_\text{int}\rangle-i\text{Tr}[({\bf \Delta}\ast {\bf G})^<]$ \cite{Aoki_2014}. 
Figure~\ref{fig_phasediagram}(c) plots the change in the entropy per site for the paramagnetic normal states, relative to the value at $T=10$. At $T\lesssim 0.075$ we see the appearance of a discontinuity associated with the high-spin/low-spin transition, where the entropy per site changes by approximately $\ln(3)$. On the bottom plane, we also show the constant entropy contours, which on the low-spin side of the spin-state crossover  converge towards the low-temperature phase transition line. Two of these contours are indicated as dashed gray lines in Fig.~\ref{fig_phasediagram}(a). 

Now we consider crystal-field quenches in the vicinity of the spin-state transition. We start at $\Delta_\text{cf}=1.75$ and $T=0.2$ (empty gray dot in Fig.~\ref{fig_phasediagram}(a)), which is above the maximum $T_c$ of about $0.145$, and suddenly reduce $\Delta_\text{cf}$. To allow a symmetry breaking, we apply a constant small seed field of $10^{-3}(c^\dagger_{1\uparrow}c_{2\downarrow}-c^\dagger_{1\downarrow}c_{2\uparrow} + \text{h.c.})$. If instead of a quench, we would perform a slow adiabatic ramp, the system would stay in equilibrium, follow the gray dashed line, 
and enter the symmetry-broken phase with $\text{Im}\phi^Y\ne 0$ around $\Delta_\text{cf}\approx 1.48$. After a quench, however, it is only possible to determine the point in the phase diagram which will be reached after thermalization (see full gray dots in Fig.~\ref{fig_phasediagram}(a)), while the trajectory from the initial to this final state involves nonthermal states. 

Figure~\ref{fig_phiy}(a) shows the time evolution of $\text{Re}\phi^Y$ and $\text{Im}\phi^Y$ after a quench to $\Delta_\text{cf}=1.4$ (blue lines) and $1.3$ (black lines), while the arrows indicate the values of $\text{Im}\phi^Y$ that will be reached after thermalization. These reference values are obtained by calculating $E_\text{tot}$ after the quench and searching for the thermal system with the post-quench value of $\Delta_\text{cf}$ and the same $E_\text{tot}$. After the quench to $\Delta_\text{cf}=1.4$, the system is expected to thermalize in the EI, at a temperature substantially below the initial $T=0.2$, which is the result of entropy cooling, i.e., the electronic system cools down because the entropy in the spin sector increases after the quench. This is qualitatively similar to the adiabatic case. 

More surprising is the result after the quench to $\Delta_\text{cf}=1.3$, where the energy injected by the quench dominates the cooling effect and results in a thermalization at $T=0.238$, far above the equilibrium EI phase (see gray dot in Fig.~\ref{fig_phasediagram}(a)). Nevertheless, the complex OP grows to values comparable to the previous simulation, which shows that there occurs a symmetry breaking to a nonthermal EI with $|\phi^Y|>0$. While this transient electronic order will melt at long enough times, the DMFT results show that it lives much longer than the longest simulation time, which for typical bandwidths corresponds to $O(100)$ fs. As shown in the SM, excitonic order is not induced in quenches from high-spin states due to the lack of entropy cooling. 

It is instructive to plot the traces of the complex OPs for different values of the post-quench $\Delta_\text{cf}$,  see Fig.~\ref{fig_phiy}(b). In the same panel we also indicate by colored dots the values of the OPs reached after thermalization. For $1.35\le \Delta_\text{cf} < 1.5$ we clearly see a quench-induced symmetry breaking, with an OP that rotates around the thermal value reached in the long-time limit in an anticlockwise fashion. Between $\Delta_\text{cf}=1.35$ and $1.3$, a dynamical phase transition \cite{Eckstein2009,Schiro2010} occurs. For $\Delta_\text{cf}\le 1.3$, the thermal OP is zero, and the transient OP encircles the origin in a clockwise fashion. Still, for $\Delta_\text{cf}\approx 1.3$ the modulus of the OP reaches large values and the nonthermal EI state is long-lived. In contrast to the thermal OP with $\text{Re}\phi^Y=0$, this nonthermal OP reaches its largest values along the real axis. Since the excitonic order is related to magnetic dipoles/multipoles~\cite{Nasu_2016}, the transient order observed here implies oscillations of magnetic dipoles/multipoles and thus oscillations of the local spin susceptibility.

A clear picture of the nonequilibrium evolution can be obtained from the mean-field (MF) dynamics of a strong-coupling effective model defined in the space of the four dominant half-filled states (one low-spin state $|L\rangle$ and three high-spin states $|H_X\rangle$,$|H_Y\rangle$,$|H_Z\rangle$) \cite{Kunes_2014,Kunes_2015,Nasu_2016}.
As detailed in the SM, this model has the form    
\begin{align}
\hH_{\rm eff} &= -h_z \sum_{i} \htau_{i}^z - h_{\rm seed}\sum_i  \htau_{Y,i}^y +J_s \sum_{\langle ij\rangle} {\bf \hS}_i\cdot  {\bf \hS}_j  \label{eq:H_eff2} \\
&+ J_z \sum_{\langle ij\rangle } \htau_i^z \htau_j^z -J_x\sum_{\langle ij\rangle} \sum_\Gamma \htau^x_{\Gamma i}  \htau^x_{\Gamma j}  -J_y \sum_{\langle ij\rangle}  \sum_\Gamma \htau^y_{\Gamma i}  \htau^y_{\Gamma j},  \nonumber  
\end{align}
where $\tfrac12\htau_{\Gamma}^{x,y,z}$ are pseudospin-$\tfrac12$ operators defined in the subspace of $|H_{X}\rangle$ and $|L\rangle$, $\htau_{i}^z=\sum_{\Gamma}\htau_{\Gamma,i}^z$, $h_{\rm seed}$ is the seed field, $\hS$ is the spin-1 operator defined in the high-spin subspace, and $h_z$, $J_{x,y,z,s}$ are determined by $v$, $v'$, $U$, $J$, $I$, and $\Delta_\text{cf}$.
 The low-spin state $|L\rangle$ yields $\tau_\Gamma^z=-1$ and zero expectation values for the other $\hat \tau$ and $\hat S$, while the OP $\phi^\Gamma$ corresponds to $\langle \hat\tau_\Gamma^x+i\hat\tau_\Gamma^y\rangle$.
  Figure ~\ref{fig_phiy}(c) shows the MF dynamics of the effective model after quenches from a low-spin state above the maximum $T_c$.
 The main difference to the DMFT results is the absence of a gradually shrinking contour, which is attributed to the lack of  thermalization.
 Still, the MF results yield a qualitatively similar dynamics, including the dynamical phase transition. 
As shown in the SM, the MF dynamics is simple, because $\tau_X^{x,y},\tau_Z^{x,y}$ and $S^{x,y,z}$ stay zero and the occupations of $|H_X\rangle$ and $|H_Z\rangle$ remain constant ($m$). Hence, one can focus on the dynamics of $ {\btau_{Y}}$, which is described by the simplified Hamiltonian 
$H'_\text{eff}=
-h_z'\sum_i \htau_{Y,i}^z  - h_{\rm seed}\sum_i  \htau_{Y,i}^y +4J_z\sum_{\langle i,j\rangle} \htau_{Y,i}^z \htau_{Y,j}^z
-J_x\sum_{\langle i,j\rangle} \htau_{Y,i}^x \tau_{Y,j}^x -J_y\sum_{\langle i,j\rangle} \htau_{Y,i}^y \htau_{Y,j}^y$,
where $-h_z'= -2h_z+2z_n(4m-1)J_z$, and $z_n$ is the number of neighboring sites. Thus, in the MF evolution, $|\btau_Y|$ is conserved.
In equilibrium, the pseudo-spin $\btau_Y$ is aligned with the pseudo-magnetic field (${\bf h}_\text{MF}$), where $h_\text{MF}^x=0$ and $-h_\text{MF}^z\gg h_\text{MF}^y>0$.
The quench of $\Delta_{\rm cf}$ decreases $-h_\text{MF}^z$ ($>0$) and the initial evolution of the OP corresponds to a pseudo-spin precession around the quenched pseudo-magnetic field.

\begin{figure}[t]
\begin{center}
\includegraphics[angle=0, width=1.0\columnwidth]{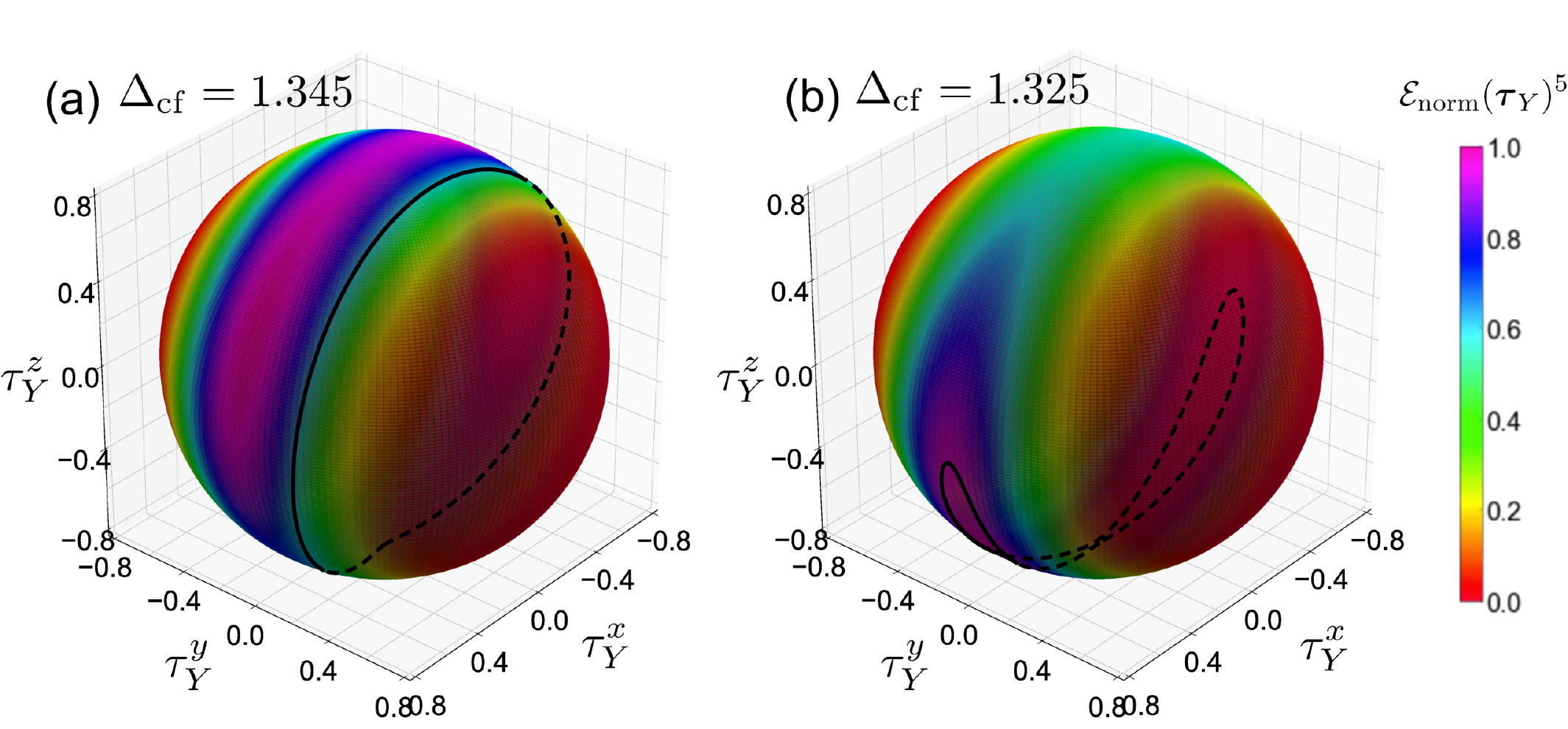}%\hfill
\caption{MF trajectories of $\btau_Y$ and the total energy $\mathcal{E}(\btau_Y)$ for quenches to the indicated values of $\Delta_{\rm cf} $ from $\Delta_{\rm cf} = 2.0,T=0.3$ ($U=6,J=1, I = \frac{\Delta_{\rm cf}}{1.5}$, $h_{\rm seed}=\sqrt{2}\cdot 10^{-3}$). Trajectories and energies are plotted on a Bloch sphere with constant $|\btau_Y|$ (determined by the initial state).  $\mathcal{E}_{\rm norm}(\btau_Y)$ is $\mathcal{E}(\btau_Y)$ normalized to the range $[0,1]$.
}
\label{fig_energy}
\end{center}
\end{figure}

The MF trajectory of $\btau_{Y}$ can be understood by considering the total energy of the effective model,  
$\mathcal{E}(\btau_Y)$, which is a function of the three angles associated with the vector $\btau_Y$. Since the energy after the quench is conserved, $\btau_Y$ follows a constant-energy contour on the Bloch sphere.
In Fig.~\ref{fig_energy}, we plot the $\btau_Y$-trajectories and the energy on both sides of the dynamical phase transition. 
In all cases, the energy minima are at $(\tau_Y^x,\tau_Y^y,\tau_Y^z)=(0,\pm A, B)$ ($A>0$) 
and before the dynamical phase transition, the $\btau_Y$-trajectories encircle this minimum.
As we approach the transition, (i) maxima in $\mathcal{E}$ emerge at  $(\tau_X^x,\tau_X^y,\tau_X^z)=(\pm A, 0,B)$. 
The transition occurs when (ii) $\mathcal{E}(0,0,|\tau_Y|)>\mathcal{E}(0,0,-|\tau_Y|)$ switches to $\mathcal{E}(0,0,|\tau_Y|)<\mathcal{E}(0,0,-|\tau_Y|)$. (i) and (ii) combined have a drastic effect on the accessible contours \cite{comment} 
and result in a switch of the pseudo-spin dynamics from a rotation around the $\tau_Y^y$ axis to a rotation around the $\tau_Y^z$ axis. The existence of the maximum (i) is a consequence of $J_x\ll J_y$ for $v\simeq v'$, while, for $v'=0$, $J_x \sim J_y$, and no enhancement of the EI is observed, see SM. We note that the MF dynamics conserves the local entropy of the initial low-spin state, and in this sense the ``entropy cooling" effect is built in, and manifests itself in a constant norm of $\btau_Y$. Due to this and the properties of the constant energy contours on the Bloch sphere, there appears a large nonthermal value of the excitonic order and a dynamical phase transition. 

\begin{figure}[t]
\begin{center}
\includegraphics[angle=-90, width=0.49\columnwidth]{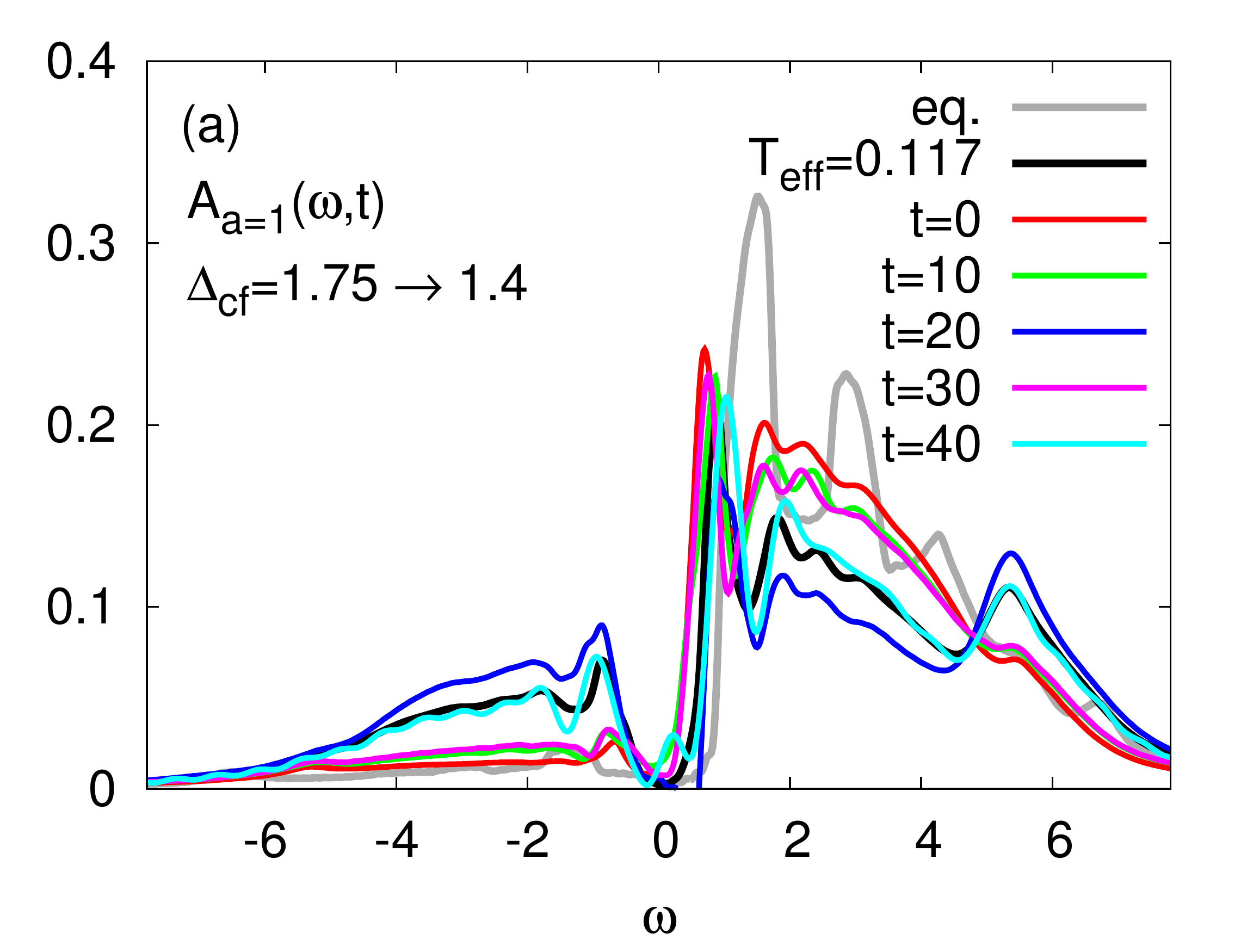} \hfill % 50 110 554 770
\includegraphics[angle=-90, width=0.49\columnwidth]{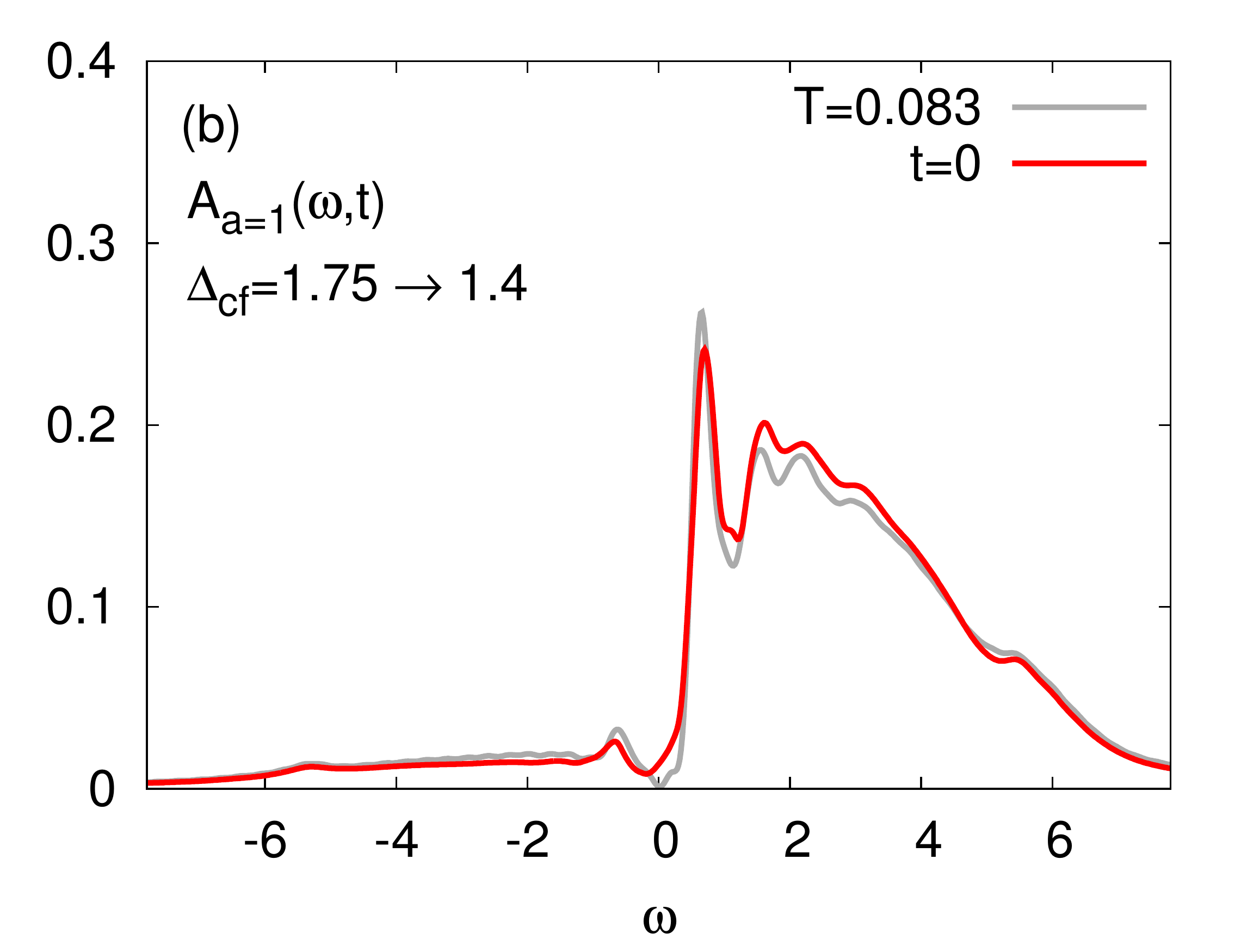} % 50 110 554 770
\includegraphics[angle=-90, width=0.49\columnwidth]{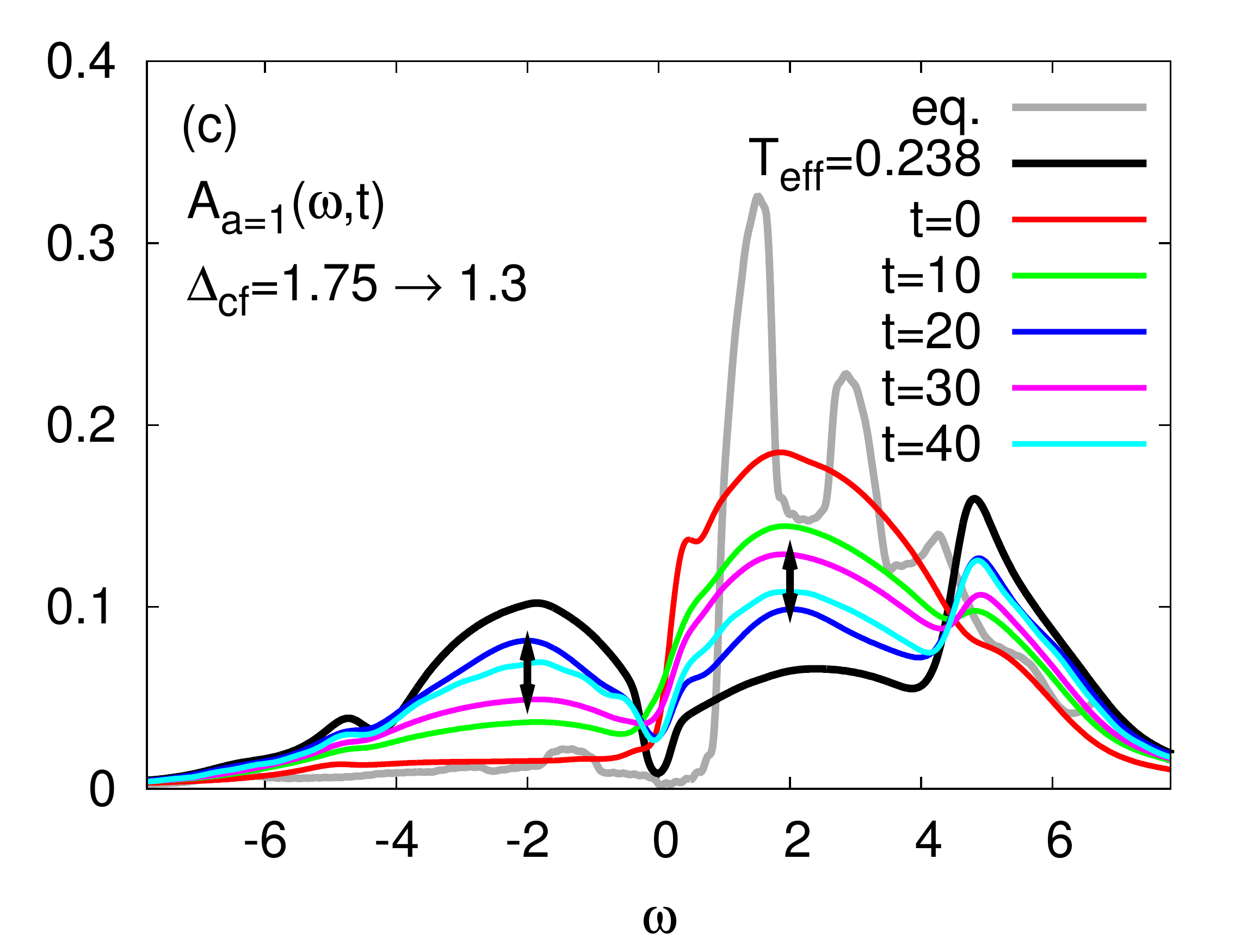} \hfill % 50 110 554 770
\includegraphics[angle=-90, width=0.49\columnwidth]{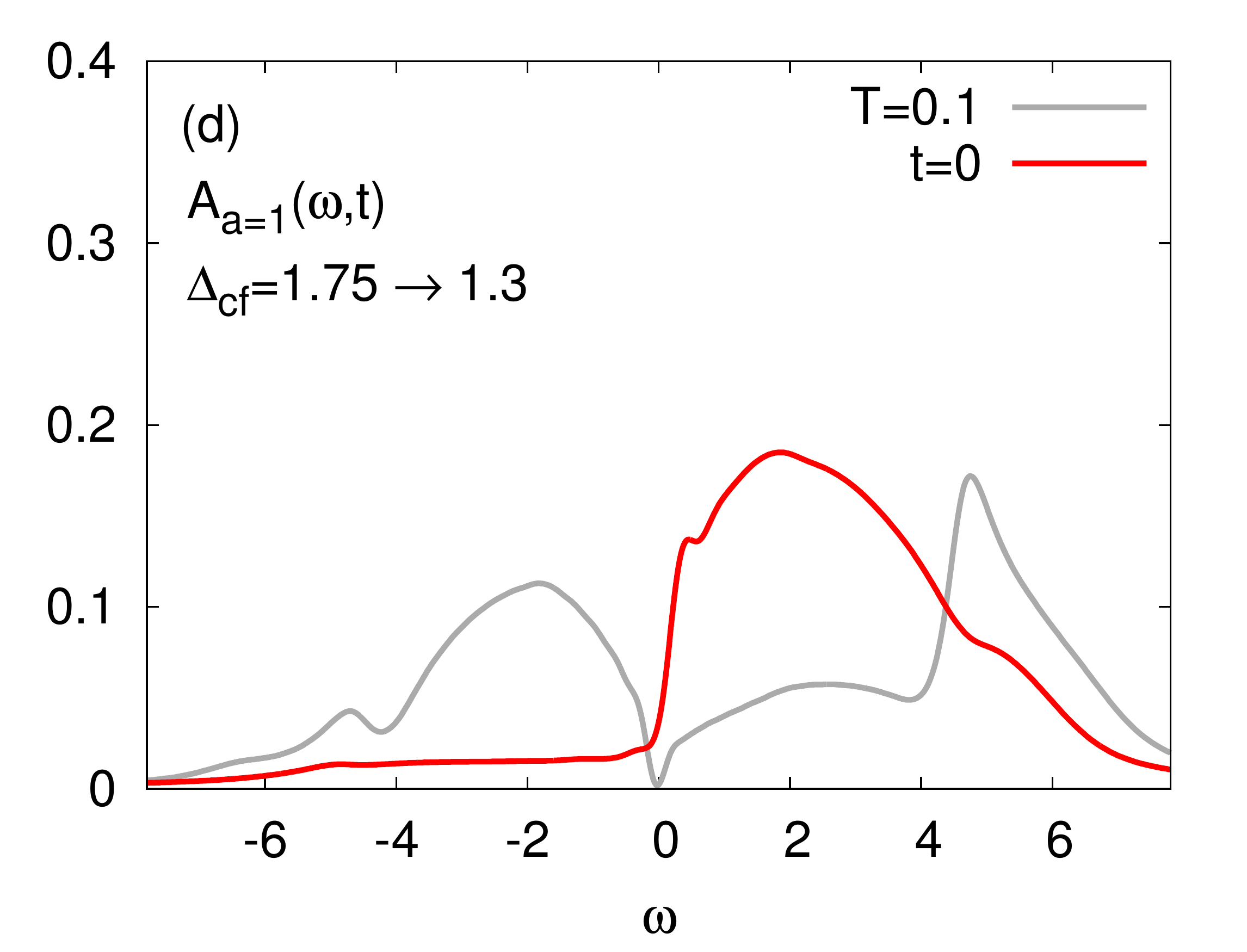} % 50 110 554 770
\caption{Time-resolved spectral function for orbital $a=1$ after a quench from $\Delta_\text{cf}=1.75$, $T=0.2$ to 
$\Delta_\text{cf}=1.4$ (a) and 1.3 (c). The gray line represents the initial equilibrium spectrum and the black line the spectrum of the thermalized state at temperature $T_\text{eff}$. The right  panels compare the nonequilibrium spectrum at $t=0_+$ to an equilibrium spectrum with $\Delta_\text{cf}=1.4$,  $T=0.083$ (b) and $\Delta_\text{cf}=1.3$,  $T=0.1$ (d).}
\label{fig_spectra_time}
\end{center}
\end{figure}

The nonthermal nature of the EI after the quench to $\Delta_\text{cf}=1.3$
is also evident in the nonequilibrium spectral function $A_{a=1,\sigma}(t,\omega)=-\frac{1}{\pi}\int_t^{t_\text{max}} dt' e^{i\omega(t-t')}G^R_{a,\sigma}(t,t')$. In Fig.~\ref{fig_spectra_time}(a,c) we show the DMFT results for quenches from $\Delta_\text{cf}=1.75$, $T=0.2$ to $\Delta_\text{cf}=1.4$ and $1.3$.
The gray curve shows the initial equilibrium spectrum, and the red curve the nonequilibrium result for $t=0_+$. These two differ, because the nonequilibrium spectrum is obtained by forward time integration. 
As can be seen in panel (b), the spectrum after the quench to $\Delta_\text{cf}=1.4$ resembles the normal-phase equilibrium spectrum for $T\approx 0.083$, demonstrating the entropy cooling effect. As time increases, we observe the opening of the excitonic gap and oscillations of the nonequilibrium spectrum around the thermalized spectrum for $T_\text{eff}=0.117$. After the quench to $\Delta_\text{cf}=1.3$, the spectrum is very different from a cold high-spin spectrum, see panel (d), and rather resembles the low-spin result shown in panel (b). As the nonthermal OP builds up we observe oscillations in the spectral function, but these are oscillations around a nonthermal type of  spectrum characterized by a large orbital polarization (black arrows in panel (c)). Also, even though the OP reaches large values in the transient state, there is no gap opening in the nonequilibrium spectrum. 

In conclusion, we showed that entropy cooling can induce a nonthermal EI and dynamical phase transition in the two-band Hubbard model. Ultra-fast photo-induced spin-state transitions have been recently reported in an $f$-electron system \cite{Mardegan_2020} and it is interesting to explore the consequences of the resulting entropy change on the nonequilibrium dynamics. An interesting topic for future studies are superconducting states without equilibrium analogue, similar to the nonthermal EI reported here.

{\it Acknowledgements}
The calculations have been performed on the Beo04 cluster at the University of Fribourg, using software libraries developed by M. Eckstein and H. Strand~\cite{Nessi}. We thank U. Staub and M. Eckstein for helpful discussions. PW acknowledges support from ERC Consolidator Grant No.~724103. YM acknowledges support from Grant-in-Aid for Scientific Research from JSPS, KAKENHI Grant Nos. JP19K23425, JP20K14412, JP20H05265, and JST CREST Grant No. JPMJCR1901.

\clearpage

\appendix

\section{SUPPLEMENTARY MATERIAL}

\section{A. Additional nonequilibrium DMFT results}

\subsection{A. 1. Equilibrium spectral functions}

\begin{figure}[ht]
\begin{center}
\includegraphics[angle=-90, width=0.8\columnwidth]{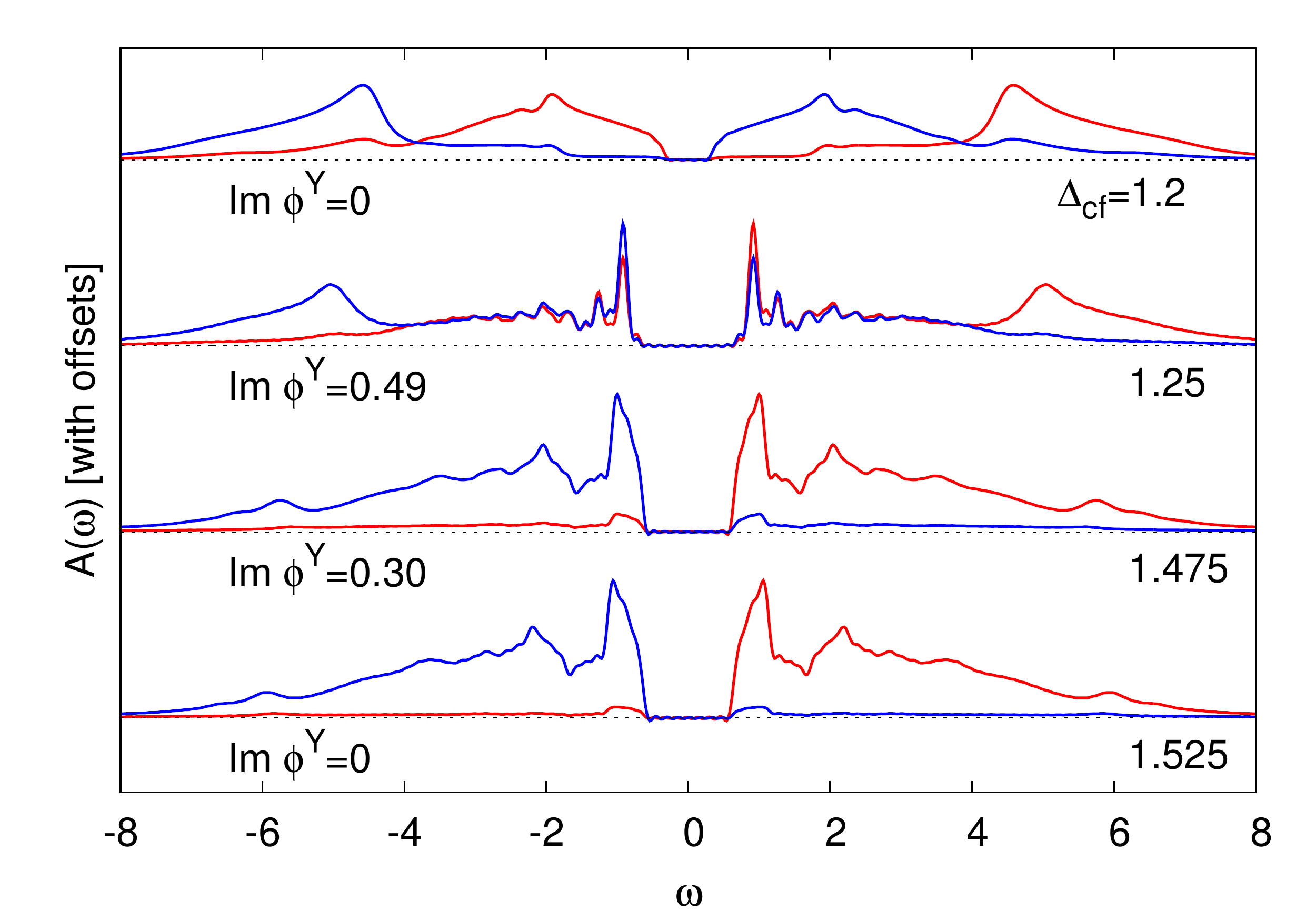} % 0 50 554 770
\caption{Equilibrium spectral functions for $G_{1,\sigma}$ (red) and $G_{2,\sigma}$ (blue) at $T=0.05$ in the vicinity of the high-spin insulator/excitonic insulator transition (top two panels) and the excitonic insulator/low-spin insulator transition (bottom two panels).}
\label{fig_spectra}
\end{center}
\end{figure}

The effect of the symmetry breaking on the spectral functions of the orbital-diagonal Green's functions at $T=0.05$ is illustrated in Fig.~\ref{fig_spectra}. The top two panels show the substantial changes ocurring in these spectra at the first order transition from the high-spin insulator to the excitonic insulator. In particular, we notice an enhancement of the gap, the appearance of sharp peaks at the gap edges, and a strongly reduced asymmetry in the spectra. As we increase the crystal field splitting inside the excitonic phase, the gap remains approximately constant, but the asymmetry in the spectra (orbital polarization) gets enhanced. The transition to the low-spin insulator, illustrated in the bottom two panels, has almost no effect on the gap and the shape of the spectral function, consistent with the continuous transition seen in Fig.~1(b) of the main text.

\subsection{A. 2. Effect of the initial temperature}

\begin{figure}[h]
\begin{center}
\includegraphics[angle=-90, width=0.8\columnwidth]{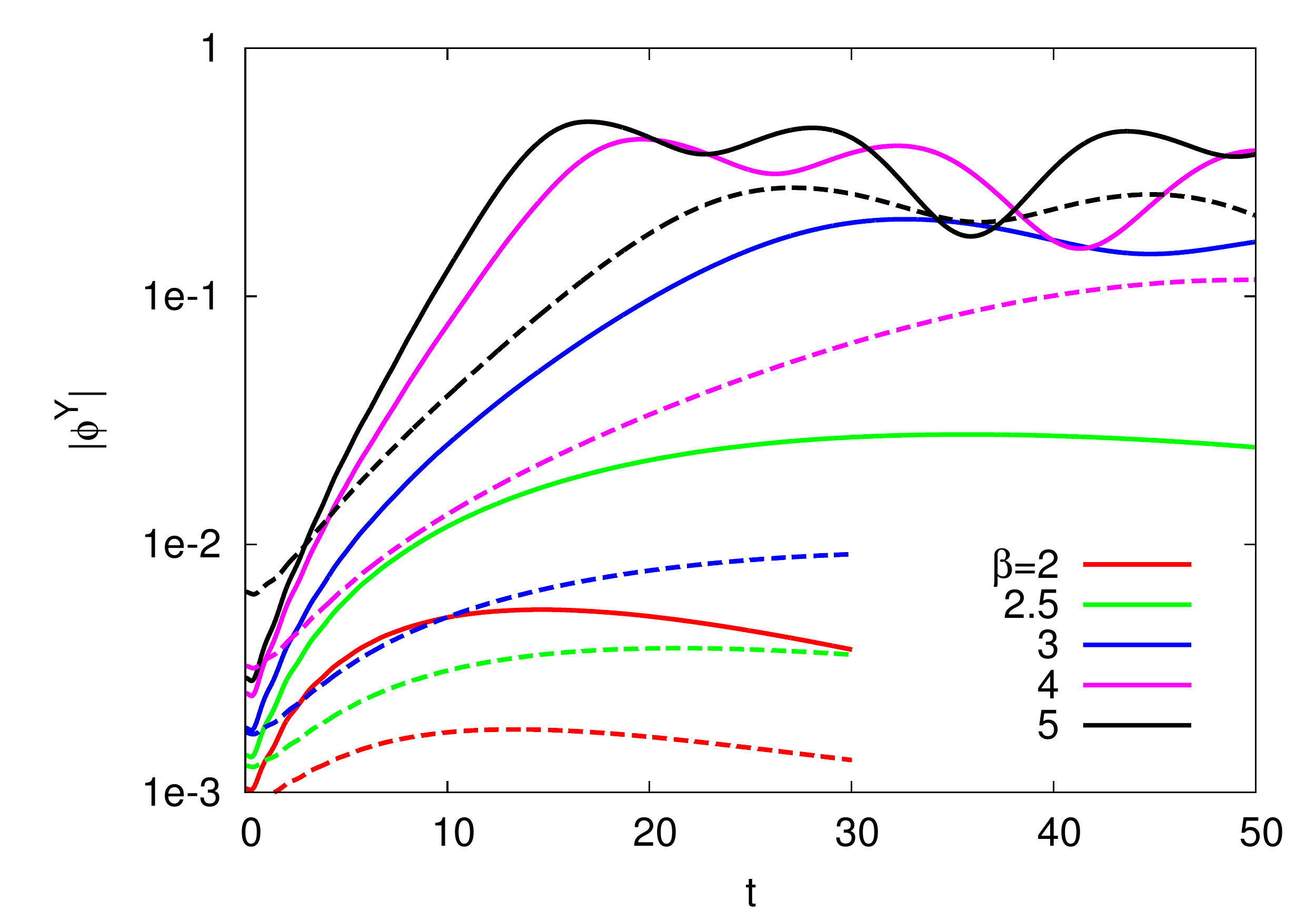} % 0 50 554 770
\caption{Modulus of the order parameter after a quench from $\Delta_\text{cf}=1.75$ (solid lines) or $\Delta_\text{cf}=1.5$ (dashed lines) to $\Delta_\text{cf}=1.35$, for indicated values of the initial inverse temperature $\beta$.}
\label{fig_beta_3.5}
\end{center}
\end{figure}

In Fig.~\ref{fig_beta_3.5} we show the evolution of $|\phi^Y|$ for quenches from $\Delta_\text{cf}=1.75$ and $1.5$ to $\Delta_\text{cf}=1.35$ and several initial inverse temperatures $\beta$. For the applied seed field of 0.001, a symmetry broken state corresponds to $|\phi^Y|\gtrsim 0.1$. Hence, in the case of the larger initial crystal field, a condensation can be induced for an initial $\beta\gtrsim 2.75$, i.e. for $T_\text{initial}\lesssim 0.4$, while in the case of the smaller crystal field, we find $\beta \gtrsim 3.5$, or $T_\text{initial} \lesssim 0.29$. Although a quench is a highly non-adiabatic protocol, this is remarkably close to the results one can deduce from the entropy contours  in Fig.~1(c) for the case of adiabatic ramps. Hence, the cooling effect after a quench is of the same order of magnitude as in the adiabatic case. The result for $\Delta_\text{cf}=1.75$ furthermore shows that condensation can be induced even if $T_\text{initial}$  is three times larger than the highest equilibrium condensation temperature ($T_\text{max}\approx 0.14$). 

\subsection{A. 3. Nonequilibrium distribution function}

\begin{figure}[t]
\begin{center}
\includegraphics[angle=-90, width=0.8\columnwidth]{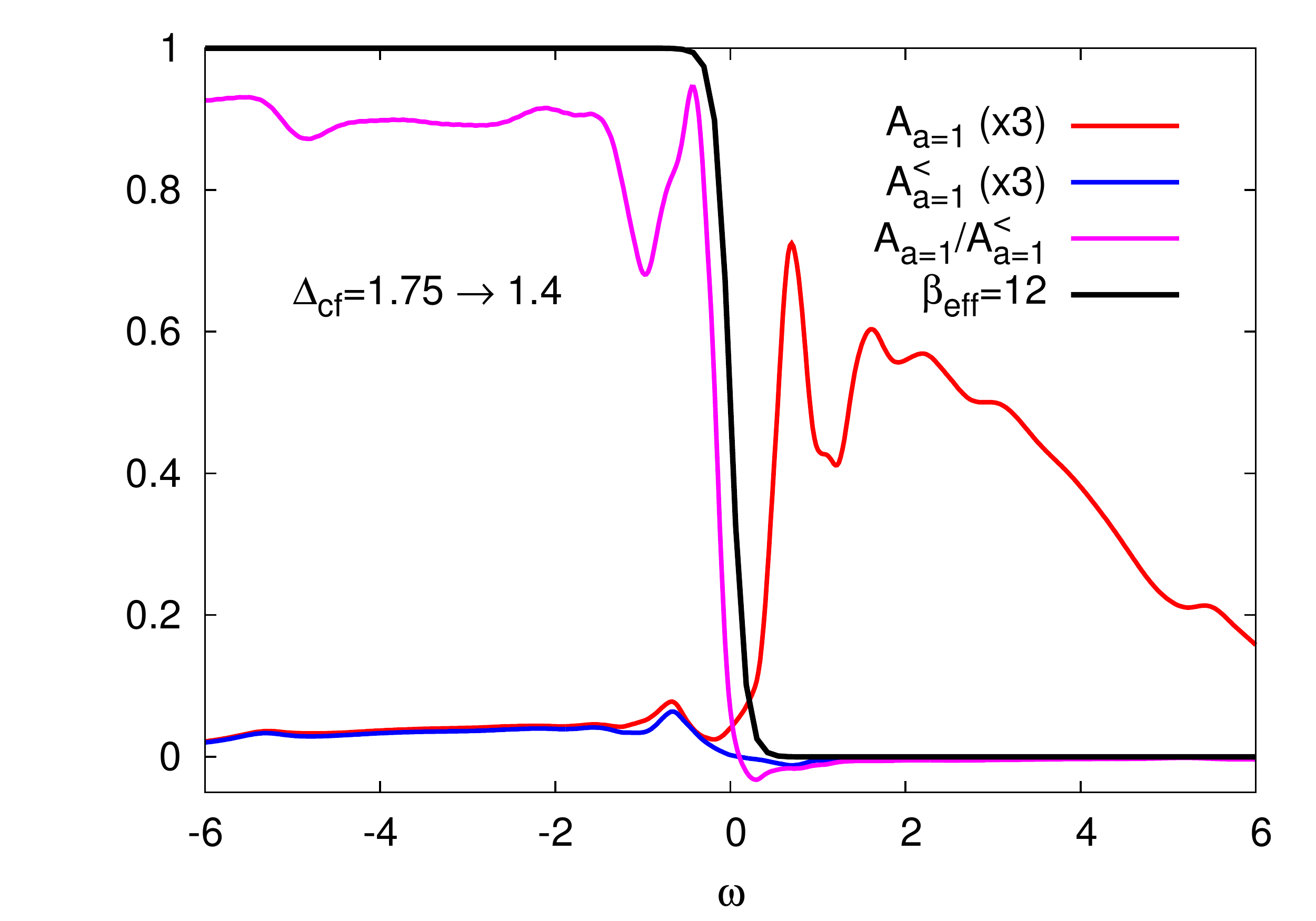}
\caption{Spectral function $A$, occupation $A^<$, and nonequilibrium distribution function $A^</A$ for orbital $a=1$ measured at $t=0_+$ after a quench from $\Delta_\text{cf}=1.75$, $\beta=5$ to $\Delta_\text{cf}=1.4$.}
\label{fig_dist}
\end{center}
\end{figure}

While the system after a quench is in a nonequilibrium state and its temperature is not defined, we can extract an effective temperature $T_\text{eff}$ (or inverse effective temperature $\beta_\text{eff}$) from the slope of the distribution function at low energies. In Fig.~\ref{fig_dist} we plot $A_{a=1}(\omega,t=0_+)$, $A^<_{a=1}(\omega,t=0_+)$ and the distribution function $f_{a=1}(\omega,t=0_+)=A^<_{a=1}(\omega,t=0_+)/A_{a=1}(\omega,t=0_+)$ obtained from the forward Fourier transformation of the nonequilibrium Green's functions at $t=0_+$. The black line shows a Fermi distribution function for inverse temperature $\beta=12$, which has approximately the same slope as $f_{a=1}(\omega,t=0_+)$. The $\beta_\text{eff}=12$ ($T_\text{eff}=0.083$) extracted from this analysis is compatible with the temperature deduced in Fig.~4(b) of the main text from the comparison of $A_{a=1}(\omega,t=0_+)$ to equilibrium spectra and confirms that the quench results in a reduction of the system's ``temperature" by more than a factor of two. 

\begin{figure}[b]
\begin{center}
\includegraphics[angle=-90, width=0.8\columnwidth]{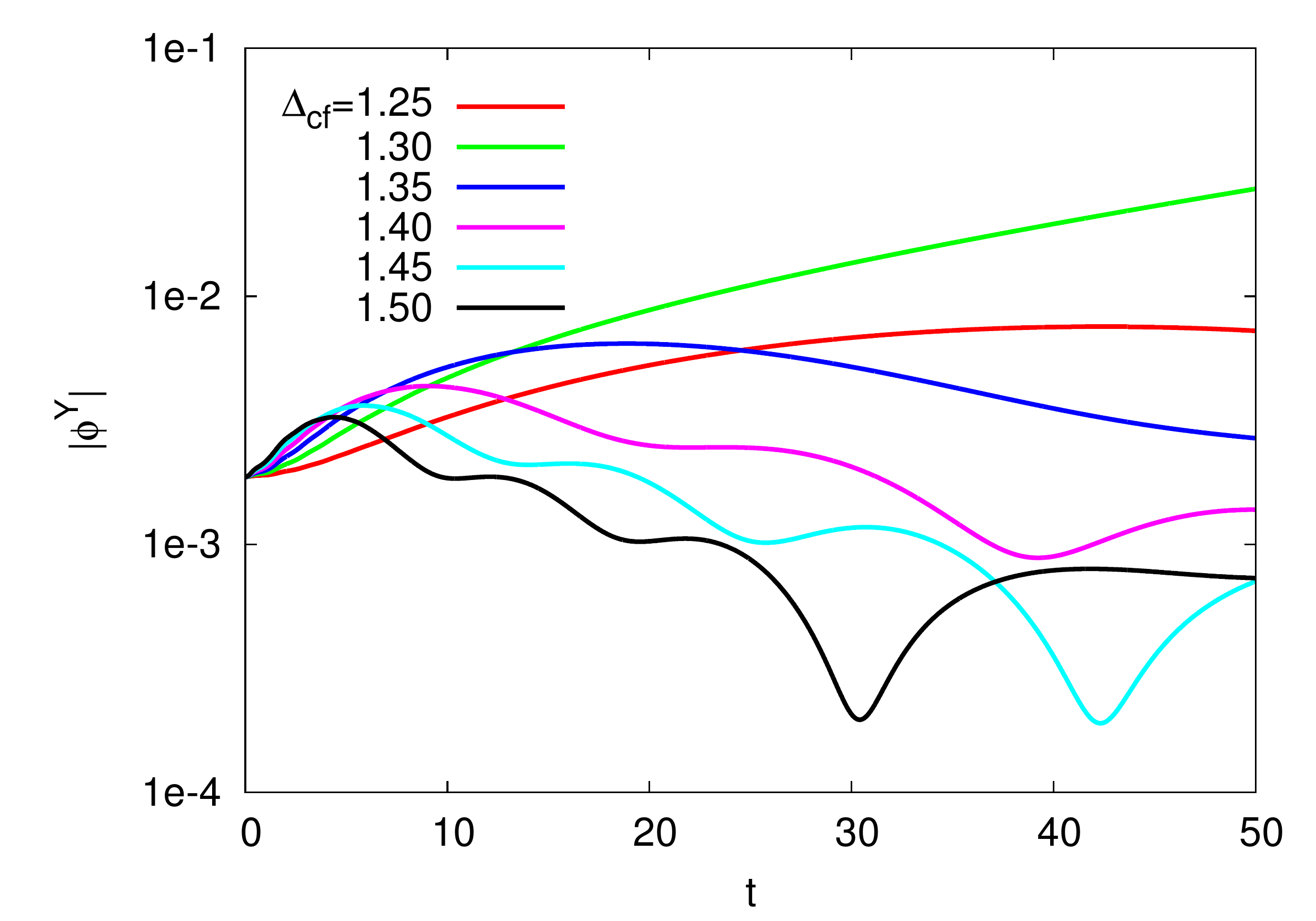} % 0 50 554 770
\caption{Modulus of the order parameter after a quench from $\Delta_\text{cf}=1.2$, $T=0.2$ to the indicated larger values of $\Delta_\text{cf}$.}
\label{fig_24}
\end{center}
\end{figure}

\subsection{A. 4. Quenches from the high-spin state}

Figure~\ref{fig_24} plots $|\phi^Y|$ for quenches which increase the crystal field splitting, starting in the high-spin region of the phasediagram ($\Delta_\text{cf}=1.2$, $T=0.2$). We again apply a seed field of 0.001, so that order parameters below $\sim 0.1$ should be considered as states with large excitonic susceptibility, but no long-range order.  
Consistent with the shape of the isentropy contours in Fig.~1(c) of the main text, we find that quenches to $\Delta_\text{cf}\le 1.3$ lead to an enhancement of the susceptibility, while those to larger crystal fields result in a heating of the system and a suppression of the seed-induced order parameter. 
Thus, we find no quench-induced symmetry breaking, in contrast to the calculations starting from low-spin states reported in the main text and above. The difference originates from a less efficient entropy cooling. 
More specifically, for quenches from a high-spin state, the number of relevant local states increases from 3 to 4. On the other hand, for quenches from low-spin states, this number increases from 1 to 4 (near the spin-state crossover) or 1 to 3 (deeper inside the high-spin region).

\subsection{A. 5. Quenches in the model with $v'=0$}

\begin{figure}[t]
\begin{center}
\includegraphics[angle=-90, width=0.8\columnwidth]{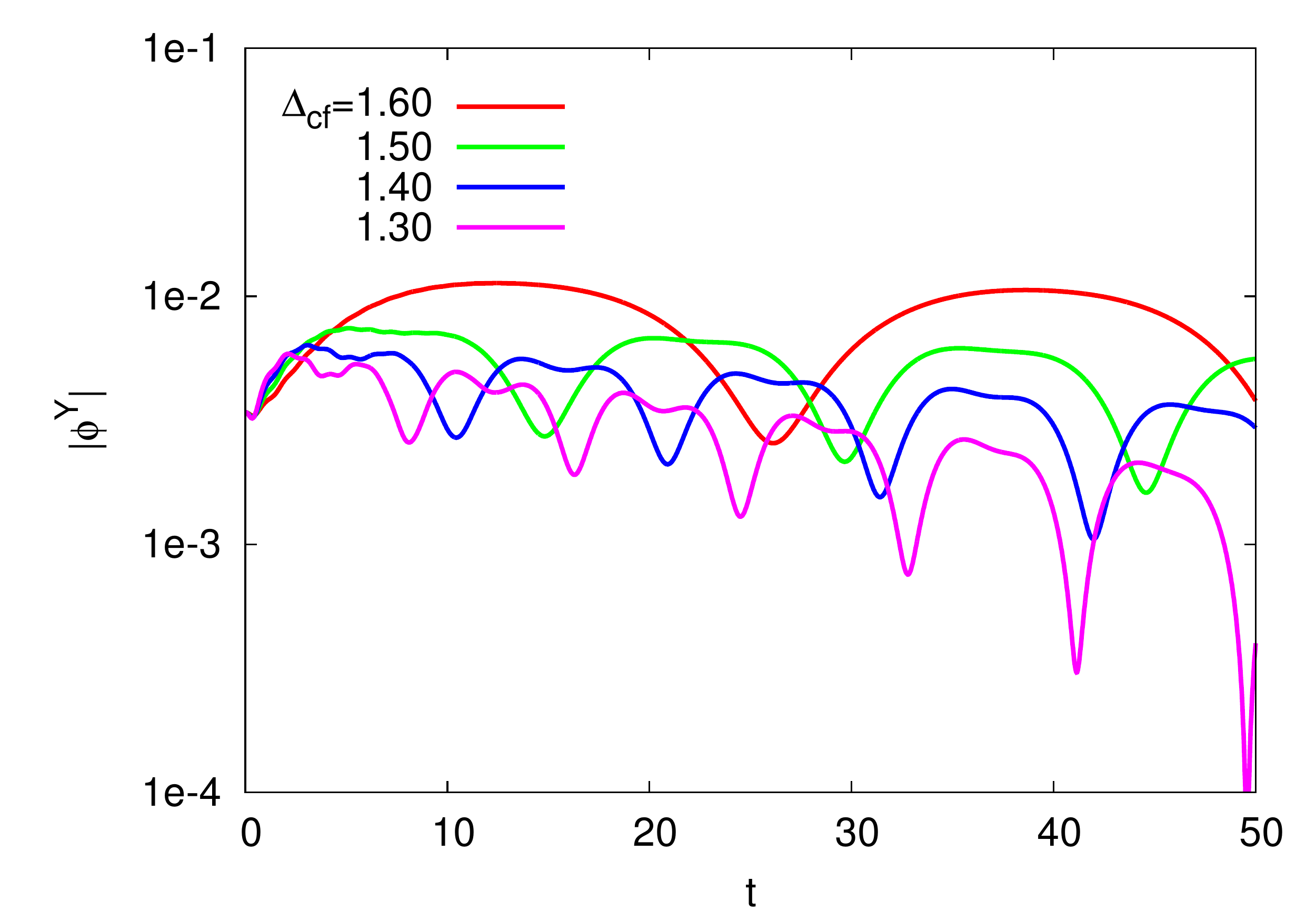}
\caption{Modulus of the order parameter after a quench from $\Delta_\text{cf}=1.85$, $T=0.2$ to the indicated larger values of $\Delta_\text{cf}$ in the model with $v=1$ and $v'=0$.}
\label{fig_tp0}
\end{center}
\end{figure}

If the cross-hopping $v'$ is set to zero, the high-spin/low-spin crossover shifts to larger values of the crystal field splitting ($\Delta_\text{cf}\approx 1.5$). While the equilibrium phase diagram still exhibits an excitonic insulator phase in the vicinity of the transition, it is no longer easy to induce such an order after a quench, at least on the time scales accessible by the DMFT simulations. This is related to the fact that without $v'$ it is very difficult for the system to change the spin state. 
(Only the pair hopping term can reshuffle charge between the orbitals, and this reshuffling by itself cannot produce high-spin states.) In Fig.~\ref{fig_tp0}, we plot the time evolution of $|\phi^Y|$ for crystal field quenches from $\Delta_\text{cf}=1.85$, $T=0.2$ to smaller values of the crystal field, in the vicinity of the high-spin/low-spin crossover. The results suggest a reduction of the order parameter up to $t=50$.

%%%%%%%%%%%%%%%%%%%%%%%%%%%%%%%%%%%%%%%%%%%%

\section{B. Effective model for the Strong coupling regime}
The original Hamiltonian defined in the main text can be expressed as follows 
\eqq{
\hH_{\rm t} = & \sum_{\langle i,j\rangle,\sigma} [v_0 \hc^\dagger_{i1\sigma} \hc_{j1\sigma} -v_0 \hc^\dagger_{i2\sigma} \hc_{j2\sigma}  + \text{h.c.}] \nonumber\\
& +  \sum_{\langle i,j\rangle,\sigma} [v_0' \hc^\dagger_{i1\sigma} \hc_{j2\sigma} + v_0' \hc^\dagger_{i2\sigma} \hc_{j1\sigma} + \text{h.c.}],\\
 \hH_{\rm chem} &= -\mu \sum_{i\sigma} ( \hn_{i1\sigma}+\hn_{i2\sigma}),
 }
 \eqq{
\hH_{\rm cf} &=    \Delta_{\rm cf} \sum_{i\sigma} (\hn_{i1\sigma}-\hn_{i2\sigma}),\\
\hH_{\rm int} = & \hspace{1mm} U \sum_{i,c=1,2} \hn_{ic\uparrow} \hn_{ic\downarrow}  + U'\sum_{i,\sigma}  \hn_{i1\sigma}\hn_{i2\bar{\sigma}} \\
&\hspace{-8mm}+ (U'-J)\sum_{i,\sigma} \hn_{i1\sigma}\hn_{i2\sigma}, \nonumber \\
 &\hspace{-8mm}+ J\sum_{i,\sigma} \hc^\dagger_{i1\sigma}\hc^\dagger_{i2\bar{\sigma}}\hc_{i1\bar{\sigma}}\hc_{i2\sigma}  + I \sum_i [\hc^\dagger_{i1\uparrow}\hc^\dagger_{i1\downarrow}\hc_{i2\downarrow} \hc_{i2\uparrow} + \text{h.c.}],\nonumber
}
with the subscript $1,2$ denoting the two orbitals, $U'=U-2J$, $v_0= v/\sqrt{z_n}$,  $v'_0= v'/\sqrt{z_n}$, and $z_n$ the number of neighboring sites.

The effective strong coupling model is defined in the space of the four local states that have similar energies in the atomic limit near the high-spin/low-spin transition \cite{Kunes_2014,Kunes_2015,Nasu_2016},
\eqq{
&|L\rangle \equiv   (f\hc^\dagger_{2\uparrow}\hc^\dagger_{2\downarrow} - g\hc^\dagger_{1\uparrow}\hc^\dagger_{1\downarrow}) |{\rm vac}\rangle,\;\; \;
|H_1\rangle\equiv \hc^\dagger_{1\uparrow} \hc_{2\uparrow}^\dagger |{\rm vac}\rangle , \nonumber\\
&|H_0\rangle\equiv \frac{1}{\sqrt{2}} (\hc^\dagger_{1\uparrow} \hc_{2\downarrow}^\dagger + \hc^\dagger_{1\downarrow} \hc_{2\uparrow}^\dagger)  |{\rm vac}\rangle ,\;\;\;
|H_{-1}\rangle \equiv  \hc^\dagger_{1\downarrow} \hc_{2\downarrow}^\dagger |{\rm vac}\rangle ,
}
with $f=1/\sqrt{1+(\Delta'-\Delta)^2/I^2}$, $g=\sqrt{1-f^2}$, $\Delta' = \sqrt{\Delta^2+I^2}$ and $\Delta=\Delta_{\rm cf}/2$.

To express the effective model, we introduce Schwinger-like bosons defined for each local state as 
\eqq{
&|L\rangle =\hh^\dagger|\Omega\rangle,\;\; \; |H_1\rangle =\hd^\dagger_1|\Omega\rangle, \nonumber\\
&|H_0\rangle =\hd^\dagger_0|\Omega\rangle,\;\;\;|H_{-1}\rangle =\hd^\dagger_{-1}|\Omega\rangle.
}
Here, $|\Omega\rangle$ is the new vacuum state for the bosons.
In the Fock space defined by the Schwinger-like bosons, the physical states are required to obey the local constraint 
\eqq{
\hh_i^\dagger \hh_i + \sum_s \hd_{is}^\dagger \hd_{is} = 1. \label{eq:physical}
}
In addition, we introduce a new set of Bosons as 
\eqq{
{\bf \hd}^\dagger =
\begin{bmatrix}
\hd^\dagger_X \\
\hd^\dagger_Y \\ 
\hd^\dagger_Z
\end{bmatrix}
\equiv 
\frac{1}{\sqrt{2}}
\begin{bmatrix}
\hd^\dagger_{-1} - \hd^\dagger_1 \\
i(\hd^\dagger_{-1}  + \hd^\dagger_1 ) \\
\sqrt{2} \hd^\dagger_0.
\end{bmatrix}
}
and the relevant operators  
\eqq{
\hn_i = \sum_s \hd^\dagger_{is} \hd_{is},\;\;\; {\bf \hS}_i = -i {\bf \hd}_i^\dagger  \times  {\bf \hd}_i.
}

The effective Hamiltonian obtained by second order perturbation in the hopping parameters can be expressed as  \cite{Kunes_2014,Kunes_2015}
\eqq{
\hH_{\rm eff} &= \epsilon \sum_i \hn_i + K_\perp \sum_{\langle ij\rangle} ({\bf \hd}_i^\dagger \cdot {\bf \hd}_j  \hh^\dagger_j \hh_i + h.c.) + K_\parallel \sum_{\langle ij\rangle} \hn_i \hn_j  
\nonumber \\
&- K_1 \sum_{\langle ij\rangle} ({\bf \hd}_i^\dagger \cdot {\bf \hd}^\dagger_j  \hh_j \hh_i + h.c.) + K_0 \sum_{\langle ij\rangle} {\bf \hS}_i\cdot  {\bf \hS}_j.  \label{eq:H_eff}
}
The expressions for the effective parameters are given in Ref.~\onlinecite{Kunes_2014} and for the model considered here are summarized in Table~\ref{tab:coeff}. 

\begin{table*}[htb]
  \begin{tabular}{|l|c|c||c|} \hline
    $\;\; \epsilon \;\;$ & %z
    \begin{tabular}{c}
    $\Delta'-3J +2 v^2\Bigl[\frac{I^2}{\Delta'^2(U-5J+2\Delta')}-\frac{I^2}{2\Delta' (\Delta'+\Delta)(U-2J+\Delta'+\Delta)}-\frac{\Delta' + \Delta}{2\Delta'(U-2J + \Delta' -\Delta)}\Bigl]$ \\
     $\;\;\;\;\;\;\;\;+ v'^2\Bigl[\frac{I^4}{\Delta'^2 (\Delta'+\Delta)^2 (U-5J+2\Delta' + \Delta)}-\frac{2}{U-2J+\Delta'} + \frac{(\Delta' + \Delta)^2}{\Delta'^2(U-5J + 2\Delta' -\Delta)}\Bigl]$ 
    \end{tabular} \\ \hline 
    
        $\;\; K_\parallel \;\;$ & 
    \begin{tabular}{c}
    $ \frac{2 v^2}{z_n}\Bigl[-\frac{I^2}{\Delta'^2(U-5J+2\Delta')}-\frac{I^2}{(U+J)\Delta' (\Delta'+\Delta)}+\frac{I^2}{\Delta' (\Delta'+\Delta)(U-2J+\Delta'+\Delta)}
    +\frac{\Delta' (U+J-\Delta) + \Delta(3J+\Delta)}{(U+J)\Delta'(U-2J + \Delta' -\Delta)}\Bigl]$ \\
    
     $+ \frac{ v'^2}{z_n}\Bigl[-\frac{I^4}{\Delta'^2 (\Delta'+\Delta)^2 (U-5J+2\Delta' + \Delta)}+\frac{4}{U-2J+\Delta'} - \frac{(\Delta' + \Delta)^2}{\Delta'^2(U-5J + 2\Delta' -\Delta)} -\frac{2(U+J)}{(U+J)^2-\Delta^2}\Bigl]$ 
    \end{tabular} \\ \hline
    
        $\;\; K_\perp \;\;$ & 
    \begin{tabular}{c}
    $- \frac{v^2}{z_n}\Bigl[\frac{I^2}{\Delta' (\Delta'+\Delta)(U-2J+\Delta'+\Delta)}+\frac{\Delta' + \Delta}{\Delta'(U-2J + \Delta' -\Delta)}  \Bigl]
    + \frac{ v'^2}{z_n} \frac{2I}{\Delta' (U-2J+\Delta')}$ 
    \end{tabular} \\ \hline
    
          $\;\; K_0 \;\;$ & 
    \begin{tabular}{c}
    $2\frac{v^2}{z_n}\frac{1}{U+J} + 2\frac{v'^2}{z_n} \frac{U+J}{(U+J)^2-\Delta^2}$ 
    \end{tabular} \\ \hline
    
     $\;\; K_1 \;\;$ & 
    \begin{tabular}{c}
    $\frac{v^2}{z_n}\frac{2I(U-2J+\Delta')}{(U+J)\Delta'(U-5J+2\Delta')} - \frac{v'^2}{z_n}
    \Bigl[ \frac{I^2(U-2J+\Delta'+\Delta)}{\Delta'(\Delta'+\Delta)(U+J+\Delta)(U-5J+2\Delta'+\Delta)} + \frac{(\Delta'+\Delta)(U-2J + \Delta'-\Delta)}{\Delta'(U+J-\Delta)(U-5J+2\Delta'-\Delta)}\Bigl]$ 
    \end{tabular} \\ \hline
    
  \end{tabular}
  \caption{The coefficients of the effective model from Ref.~\onlinecite{Kunes_2014}. Here, $\Delta = \Delta_{\rm cf}/2$. }
  \label{tab:coeff}
\end{table*}

In the subspace defined by $|L\rangle$ and $|H\rangle$, the operators for the excitonic order parameters $\hat{\phi}^\Gamma = \sum_{\sigma\sigma'}\hc^\dagger_{1\sigma}\hc_{2\sigma'} \sigma^\Gamma_{\sigma\sigma'}$ can be expressed as 
\eqq{
\hat{\phi}^\Gamma = \sqrt{2} f \hd^\dagger_\Gamma\hh + \sqrt{2} g \hh^\dagger \hd_\Gamma .
}
In practice $\hat{\phi}^\Gamma \simeq \sqrt{2} \hd^\dagger_\Gamma\hh$, since we are interested in the case $\Delta\simeq 3J\simeq 3I$.

This Hamiltonian can also be rewritten in terms of pseudo-spins \cite{Nasu_2016}. 
To this end we introduce the pseudo-spin operators as 
\eqq{
\htau^x_\Gamma = \hd_\Gamma^\dagger \hh + \hh^\dagger \hd_\Gamma,\;\;  \htau^y_\Gamma = -i \hd_\Gamma^\dagger \hh + i \hh^\dagger \hd_\Gamma,\;\; \htau^z_\Gamma = \hd_\Gamma^\dagger \hd_\Gamma - \hh^\dagger  \hh.
}
We note that $\frac{\hbtau_\Gamma}{2} $ is a spin-$\frac{1}{2}$ operator in the subspace defined by $|H_\Gamma\rangle$ $(\equiv \hd^\dagger_\Gamma |\Omega\rangle)$ and $|L\rangle$.
With these operators, the effective Hamiltonian can be expressed as 
\eqq{
\hH_{\rm eff} = & -h_z \sum_i \htau_i^z + J_z \sum_{\langle ij\rangle } \htau_i^z \htau_j^z + J_s \sum_{\langle ij\rangle} {\bf \hS}_i\cdot  {\bf \hS}_j  \nonumber \\
&-J_x \sum_{\langle ij\rangle} \sum_\Gamma \htau^x_{\Gamma i}  \htau^x_{\Gamma j}  -J_y \sum_{\langle ij\rangle}  \sum_\Gamma \htau^y_{\Gamma i}  \htau^y_{\Gamma j} , \label{eq:H_eff2}
}
where $\htau_i^z =\sum_\Gamma \htau_{\Gamma i}^z $ and 
\eqq{
h_z = -\Bigl[\frac{\epsilon}{4} + \frac{3z_n}{16} K_\parallel \Bigl],\;\;   J_z= \frac{K_\parallel}{16},\;\;    J_s= K_0,\nonumber \\
J_x = -\frac{1}{2}(K_\perp - K_1),  \;\;   J_y = -\frac{1}{2}(K_\perp + K_1).
}
Nonzero values of $\tau_\Gamma^x$ or $\tau_\Gamma^y$ indicate excitonic order.
More specifically, $\tau_\Gamma^x$ corresponds to the real part of $\phi^\Gamma$, while $\tau_\Gamma^y$ corresponds to the imaginary part of $\phi^\Gamma$.
On the other hand, $\tau_\Gamma^z$ measures, roughly, the orbital polarization.

\section{C. Mean-field theory: Formulation}
Here we consider the equilibrium phase diagram and the dynamics within the (static) mean-field (MF) theory, which is familiar from spin systems.
It can be simply formulated by renaming the operators as
\begin{align}
[\hD^\dagger_0,\hD^\dagger_1,\hD^\dagger_2,\hD^\dagger_3] = [\hd^\dagger_X,\hd^\dagger_Y,\hd^\dagger_Z,\hh^\dagger]
\end{align}
and introducing the local density matrix 
\eqq{
\hrho_{\alpha\beta,i}\equiv \hD_{\beta i}^\dagger \hD_{\alpha i}.
}
Within the MF theory, the equilibrium state and the time evolution are described by the MF Hamiltonian obtained by the decoupling of the intersite terms,
e.g. ${\bf \hd}_i^\dagger \cdot {\bf \hd}_j  \hh^\dagger_j \hh_i \rightarrow {\bf \hd}_i^\dagger \hh_i  \cdot \langle {\bf \hd}_j  \hh^\dagger_j \rangle +  \langle{\bf \hd}_i^\dagger \hh_i \rangle \cdot  {\bf \hd}_j  \hh^\dagger_j $.
After this decoupling, the MF Hamiltonian can be written as the sum of local Hamiltonians, $\hH^{\rm MF}[\brho] = \sum_i \hH^{\rm MF}_i[\brho]$. Here, $[\brho]$ indicates that $\hH^{\rm MF}$ depends on the expectation value of $\hbrho$.

In equilibrium, the state of each site is described by the ensemble of eigenstates of $\hH^{\rm MF}_i[\brho]$ in the physical space \eqref{eq:physical} (there are four in total) with the usual Boltzmann weight.
We determine this state self-consistently, so that the expectation value of $\hbrho$ is the same as the $\brho$ used in $\hH^{\rm MF}_i[\brho]$.
The time evolution is described by the differential equation
\eqq{
\partial_t \brho_i(t) = -i[{\bf h}^{\rm MF}_i[\brho(t)],\brho_i(t)],
}
where $ {\bf h}^{\rm MF}_i$ is a $4 \times 4$ matrix satisfying $\hH^{\rm MF}_i= [\hD^\dagger_0,\hD^\dagger_1,\hD^\dagger_2,\hD^\dagger_3] {\bf h}^{\rm MF}_i[\hD_0,\hD_1,\hD_2,\hD_3]^T$.

\section{D. Mean-field theory: Results}
In the following, we consider the infinitely coordinated Bethe lattice as in the main text.
  %%%%%%%%%%%%%%%%%%%%%%%%%%%%%%%%%%%%%%%%%%%%%
 \begin{figure}[t]
  \centering
    \hspace{-0.cm}
    \vspace{0.0cm}
   \includegraphics[width=85mm]{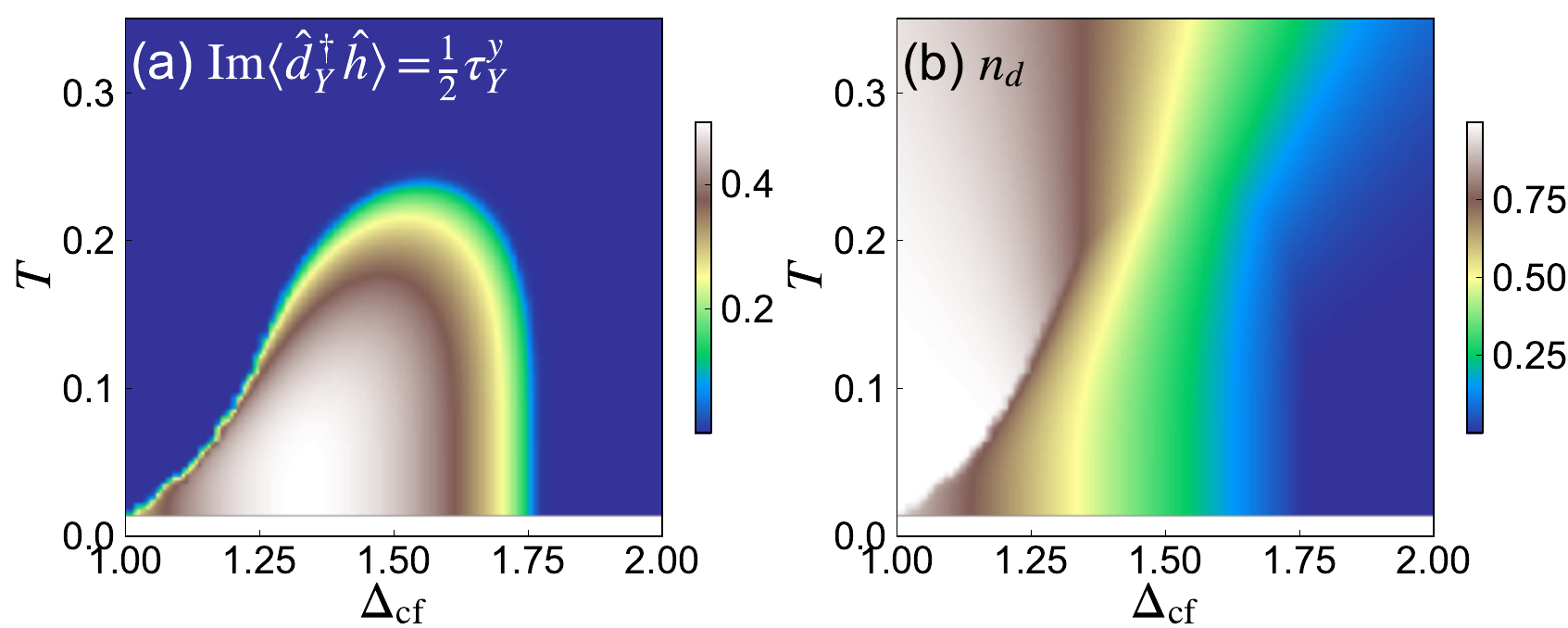} 
  \caption{(a) Excitonic order ${\rm Im}\langle \hat{d}^\dagger_Y \hh\rangle$  and (b) probability of high-spin states in the plane of crystal field $\Delta_{\rm cf}$ and temperature $T$. Here, $ v=v'=1$, $U=6$, $J=1$ and $I= \frac{\Delta_{\rm cf}}{1.5}$.}
  \label{fig:phase_v_voff}
\end{figure}
%%%%%%%%%%%%%%%%%%%%%%%%%%%%%%%%%%%%%%%%%%

  %%%%%%%%%%%%%%%%%%%%%%%%%%%%%%%%%%%%%%%%%%%%%
 \begin{figure}[h]
  \centering
    \hspace{-0.cm}
    \vspace{0.0cm}
   \includegraphics[width=85mm]{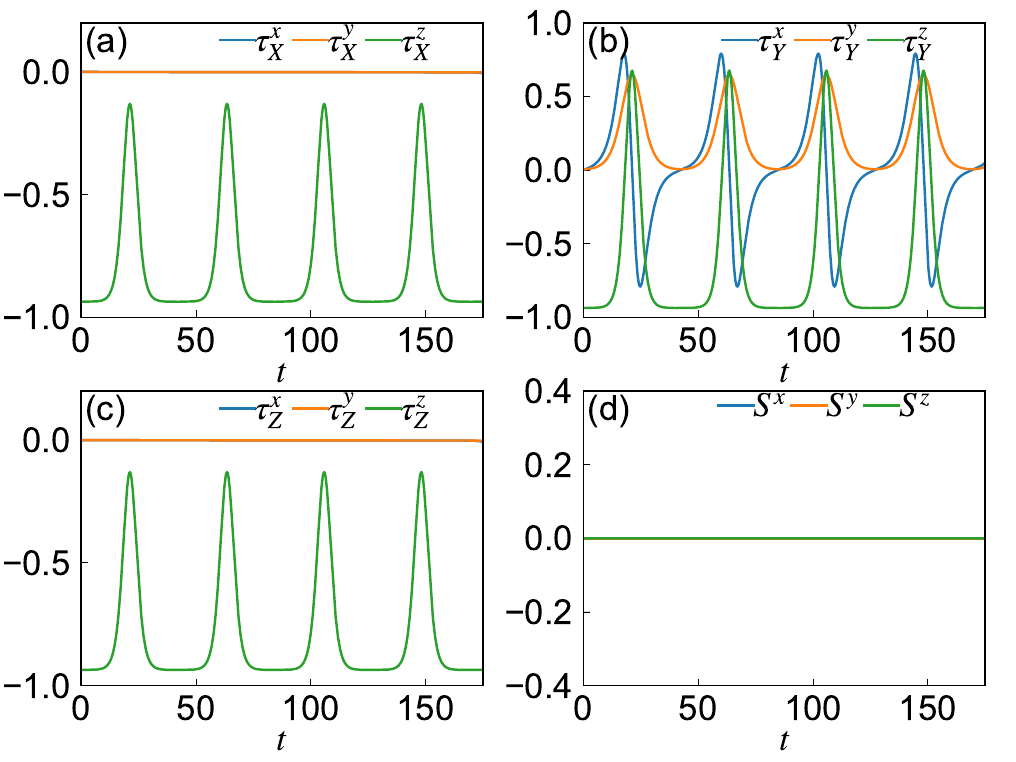} 
  \caption{Time evolution of the expectation values of the pseudo-spins (a-c) and $S=1$ spins (d) after the quench from $\Delta_{\rm cf}=2.0$ and $T=0.3$ to $\Delta_{\rm cf}=1.39$. Here, we use  $v=v'=1$, $U=6$, $J=1$, and $I=\frac{ \Delta_{\rm cf}}{1.5} $ and apply $h_{\rm seed}=\sqrt{2}\cdot 10^{-3}$ to $\sum_i  \htau_{Y,i}^y$. The parameters of the effective model after the quench become $h_z=-0.0208$, $z_nJ_x=-6.16\cdot 10^{-5}$, $z_nJ_y=0.384$, $z_nJ_z=-0.00363$, $m=0.0160$ and $r=0.936$. 
    }
  \label{fig:Quench_1}
\end{figure}
%%%%%%%%%%%%%%%%%%%%%%%%%%%%%%%%%%%%%%%%%%
\subsection{D. 1. Model with $v=v'=1$}
In Fig.~\ref{fig:phase_v_voff}(a), we show the MF phase diagram of the effective model with a small seed field proportional to $\sum_{i}\htau_{Yi}^y$ around the high-spin/low-spin crossover.
Here, we fix the ratio between $I$ and $\Delta_{\rm cf}$ so that the low spin state $|L\rangle$, with which the effective model is constructed, remains unchanged. We furthermore assume that no translational symmetry is broken.
Excitonic order appears in the high-spin/low-spin crossover region, and only ${\rm Im}\langle \hat{d}^\dagger_Y \hh\rangle$ is nonzero.
(Remember that  $\tau_{Y}^y$ corresponds to ${\rm Im}\langle \hat{d}^\dagger_Y \hh\rangle$.)
Therefore the result is consistent with the DMFT analysis, although it overestimates the region of the excitonic phase.
We note that if we apply seed fields proportional to $\sum_{i}\htau_{Xi}^y$ or $\sum_{i}\htau_{Zi}^y$, the phase shows nonzero ${\rm Im}\langle \hat{d}^\dagger_X \hh\rangle$ and  ${\rm Im}\langle \hat{d}^\dagger_Z \hh\rangle$, respectively,
instead of  nonzero ${\rm Im}\langle \hat{d}^\dagger_Y \hh\rangle$, while the values are the same.
The $\htau_{\Gamma}^y$ component is favored over the $\htau_{\Gamma}^x$ component, since in this case we have $J_y \gg  |J_x|$ in the language of the effective Hamiltonian \eqref{eq:H_eff2}.
The transition between the excitonic state and the low-spin state is of second order, while that between the excitonic state and the high-spin state is first order, as in DMFT.
Figure~\ref{fig:phase_v_voff}(b) shows the probability of the high-spin states ($n_d$). 
One can see that $n_d$ remains small up to high temperatures in the low spin regime ($\Delta_{\rm cf}\gtrsim 1.5$).
This corresponds to a large weight for the low spin state, and hence to a low entropy. 

\begin{figure}[t]
  \centering
    \hspace{-0.cm}
    \vspace{0.0cm}
   \includegraphics[width=85mm]{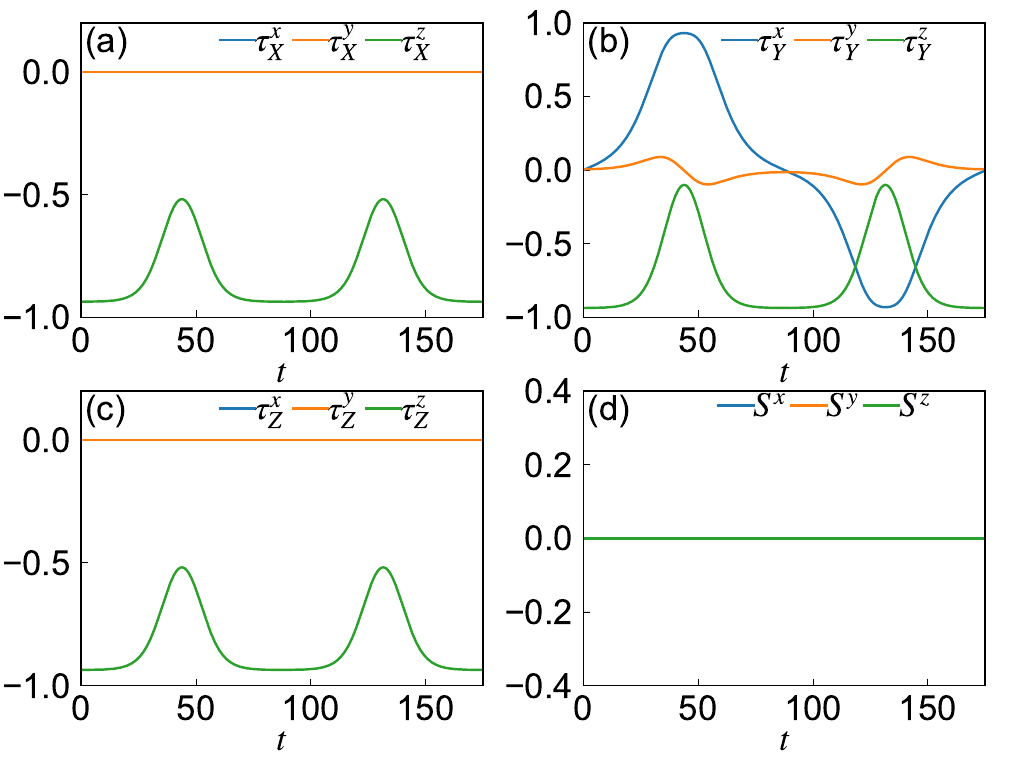} 
  \caption{Time evolution of the expectation values of the pseudo-spins (a-c) and $S=1$ spins (d) after the quench from $\Delta_{\rm cf}=2.0$ and $T=0.3$ to $\Delta_{\rm cf}=1.325$. Here, we use  $v=v'=1$, $U=6$, $J=1$, and $I=\frac{ \Delta_{\rm cf}}{1.5} $ and apply $h_{\rm seed}=\sqrt{2}\cdot 10^{-3}$ to $\sum_i  \htau_{Y,i}^y$. The parameters of the effective model after the quench become $h_z=0.00930$, $z_nJ_x=-0.000545$, $z_nJ_y=0.383$, $z_nJ_z=-0.00464$, $m=0.0160$ and $r=0.936$. }
  \label{fig:Quench_2}
\end{figure}

Now, we consider a quench from a low-spin state at a temperature which is higher than the maximum excitonic condensation temperature. 
Figure~\ref{fig:Quench_1} shows the time evolution of the expectation value of  the pseudo-spins and the $S=1$ spins after the quench from $\Delta_{\rm cf}=2.0$ and $T=0.3$ to $\Delta_{\rm cf} = 1.39$ with the seed field $h_{\rm seed} = \sqrt{2}\cdot 10^{-3}$ coupled to $\sum_i  \htau_{Y,i}^y$, i.e. $-h_{\rm seed}\sum_i  \htau_{Y,i}^y$.
The $\tau^x_Y$ and $\tau^y_Y$ components start to oscillate with large amplitudes and, in particular, the center of the $\tau^y_Y$ oscillations is $>0$.
On the other hand, Fig.~\ref{fig:Quench_2} shows the time evolution after the quench from $\Delta_{\rm cf}=2.0$ and $T=0.3$ to $\Delta_{\rm cf} = 1.325$ with the same seed field. 
Also here, the $\tau^x_Y$ and $\tau^y_Y$ components start to oscillate with large amplitude, but $\tau^y_Y$ now oscillates between positive and negative values.
These two cases represent the dynamics on the two sides of the dynamical phase transition discussed in the main text.

\begin{figure}[t]
  \centering
    \hspace{-0.cm}
    \vspace{0.0cm}
   \includegraphics[width=60mm]{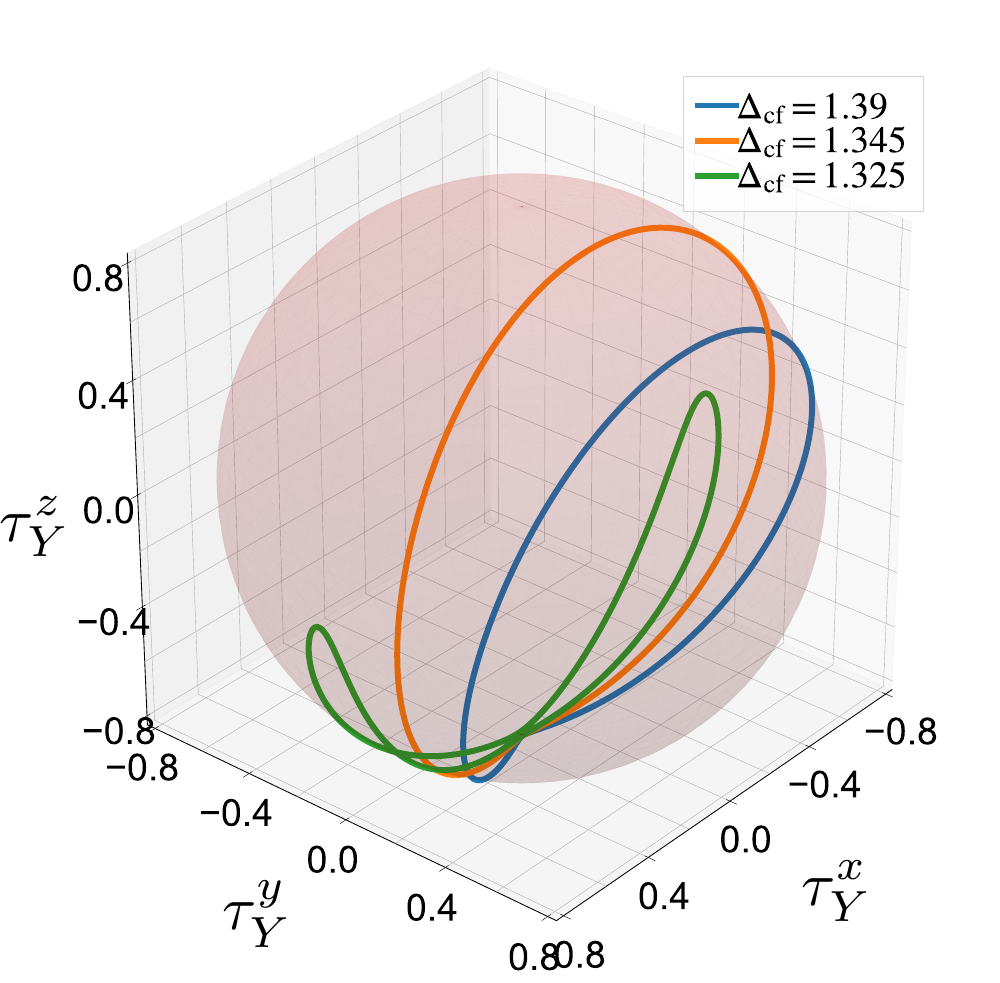} 
  \caption{Trajectories of $\btau_Y$ after quenches from $\Delta_{\rm cf}=2.0$ and $T=0.3$ to the specified values of $\Delta_{\rm cf}$. Here, we use  $v=v'=1$, $U=6$, $J=1$, and $I=\frac{ \Delta_{\rm cf}}{1.5} $ and apply $h_{\rm seed}=\sqrt{2}\cdot 10^{-3}$ to $\sum_i  \htau_{Y,i}^y$.}
  \label{fig:Quench_summary}
\end{figure}

To clarify the qualitative differences in the time evolutions, we show in Fig.~\ref{fig:Quench_summary} the trajectories of $\btau_Y$ for different $\Delta_{\rm cf}$ after the quench. 
It turns out that the norm of $\btau_Y$ remains constant and the trajectory is on a sphere, see the discussion below.
One can see that, before the dynamical phase transition ($\Delta_{\rm cf}=1.39,1.345$), the trajectories do not rotate around the $z$ axis but rotate around some axis in the $y$-$z$ plane, more specifically around $(0,A,B)$ with $A>0$.
On the other hand, after the transition, they rotate around the $z$ axis.

  %%%%%%%%%%%%%%%%%%%%%%%%%%%%%%%%%%%%%%%%%%%%%
%%%%%%%%%%%%%%%%%%%%%%%%%%%%%%%%%%%%%%%%%%

  %%%%%%%%%%%%%%%%%%%%%%%%%%%%%%%%%%%%%%%%%%%%%
%%%%%%%%%%%%%%%%%%%%%%%%%%%%%%%%%%%%%%%%%%

The dynamics of $\btau_Y$ after the quench can be understood within an effective model further simplified from Eq.~\eqref{eq:H_eff2}.
It turns out that many components ($\tau_X^{x,y},\tau_Z^{x,y}$ and  $S^{x,y,z}$) remain zero in practice, see Figs.~\ref{fig:Quench_1} and ~\ref{fig:Quench_2}.
Thus, the occupations of $|H_X\rangle$ and $|H_Z\rangle$, i.e. $n_X=\langle \hd^\dagger_X \hd_X \rangle ,n_Z=\langle \hd^\dagger_Z \hd_Z \rangle$, remain constant, and we set them to $m$ in the following.
Thanks to this, the effective Hamiltonian \eqref{eq:H_eff2} can be reduced to an effective Hamiltonian for $\hbtau_Y$ only,
 \begin{align}
H'_\text{eff}=&-h_z'\sum_i \htau_{Y,i}^z  - h_{\rm seed}\sum_i  \htau_{Y,i}^y +4J_z\sum_{\langle i,j\rangle} \htau_{Y,i}^z \htau_{Y,j}^z\nonumber\\ 
&-J_x\sum_{\langle i,j\rangle} \htau_{Y,i}^x \tau_{Y,j}^x -J_y\sum_{\langle i,j\rangle} \htau_{Y,i}^y \htau_{Y,j}^y, \label{eq:H_eff_simple}
\end{align}
where $-h_z'= -2h_z+2z_n(4m-1)J_z$, and $z_n$ is the number of neighboring sites.
Thus, within the MF dynamics, the norm of $\btau_Y$ remains constant, and we denote it by $r$.
Now the MF total energy per site, $\mathcal{E}$,  can be expressed as a function of $\btau_Y$ as 
 \begin{align}
\mathcal{E}(\btau_Y)=&-h_z' \tau_{Y}^z  - h_{\rm seed} \tau_{Y}^y +2z_nJ_z (\tau_{Y}^z)^2 \nonumber\\ 
&-\frac{z_n}{2}J_x(\tau_{Y}^x)^2 -\frac{z_n}{2}J_y\ (\tau_{Y}^y)^2. \label{eq:H_eff_energy}
\end{align}
For the purpose of illustration, we also introduce an energy $\mathcal{E}_{\rm norm}(\btau_Y)$ which is normalized to the range $[0,1]$.
This function is plotted in the main text on the sphere with radius $r$, which is determined by the initial state.
Since the energy remains constant during the time evolution, the trajectory is a constant-energy contour on this sphere. 

  %%%%%%%%%%%%%%%%%%%%%%%%%%%%%%%%%%%%%%%%%%%%%
 \begin{figure}[t]
  \centering
    \hspace{-0.cm}
    \vspace{0.0cm}
   \includegraphics[width=85mm]{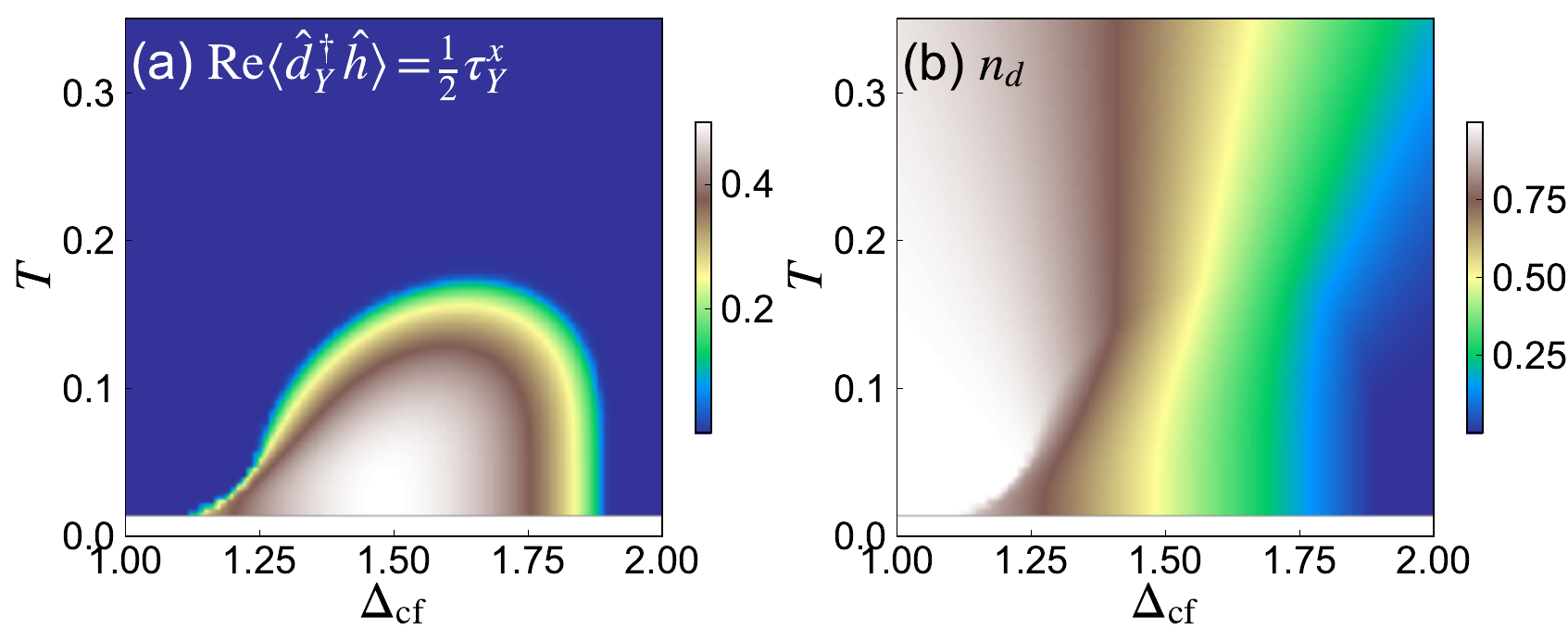} 
  \caption{(a) Excitonic order ${\rm Re}\langle \hat{d}^\dagger_Y \hh\rangle$  and (b) probability of high-spin states in the plane of the crystal field $\Delta_{\rm cf}$ and temperature $T$. Here, $v=1$, $v'=0$, $U=6$, $J=1$ and $I= \frac{\Delta_{\rm cf}}{1.5}$.}
  \label{fig:phase_voff_zero}
\end{figure}
%%%%%%%%%%%%%%%%%%%%%%%%%%%%%%%%%%%%%%%%%%

  %%%%%%%%%%%%%%%%%%%%%%%%%%%%%%%%%%%%%%%%%%%%%
 \begin{figure}[t]
  \centering
    \hspace{-0.cm}
    \vspace{0.0cm}
   \includegraphics[width=80mm]{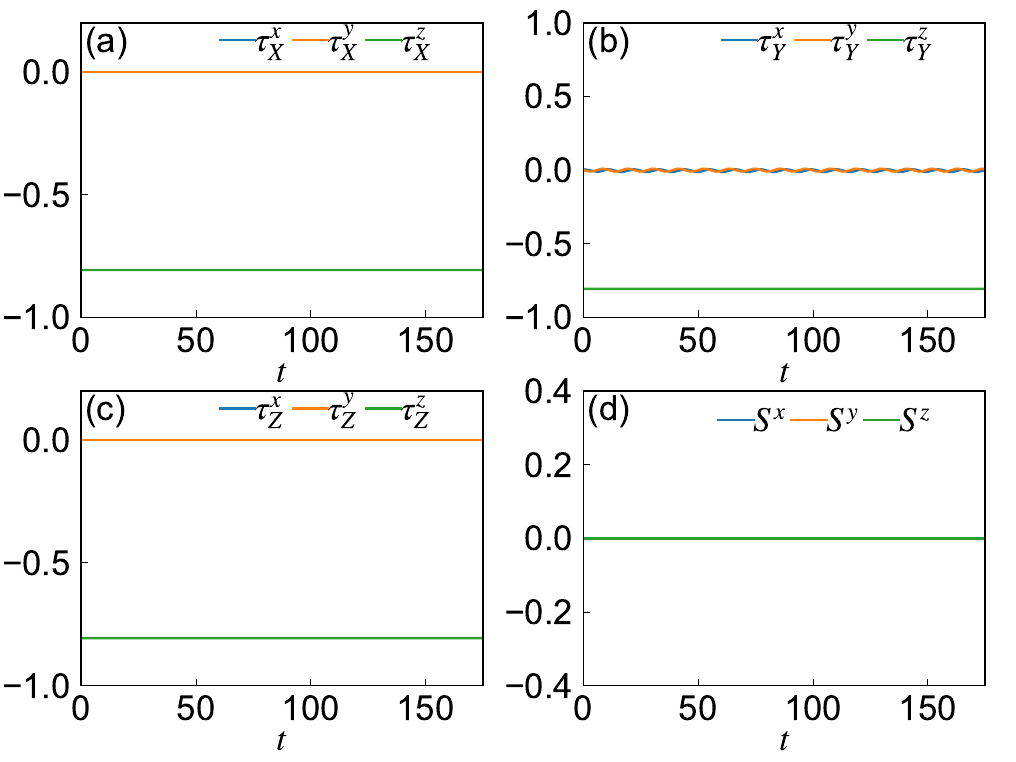} 
   \caption{Time evolution of the expectation values of the pseudo-spins (a-c) and $S=1$ spins (d) after the quench from $\Delta_{\rm cf}=2.0$ and $T=0.3$ to $\Delta_{\rm cf}=1.5$. Here, we use  $v=1$, $v'=0$, $U=6$, $J=1$, and $I=\frac{ \Delta_{\rm cf}}{1.5} $ and apply $h_{\rm seed}=\sqrt{2}\cdot 10^{-3}$ to $\sum_i  \htau_{Y,i}^x$. The parameters of the effective model after the quench become $h_z=-0.0479$, $z_nJ_x=0.281$, $z_nJ_y=0.192$, $z_nJ_z=0.0396$, $m=0.0483$, and $r=0.807$.  }
  \label{fig:Quench_voff_zero}
\end{figure}
%%%%%%%%%%%%%%%%%%%%%%%%%%%%%%%%%%%%%%%%%%

 %%%%%%%%%%%%%%%%%%%%%%%%%%%%%%%%%%%%%%%%%%%%%
 \begin{figure}[b]
  \centering
    \hspace{-0.cm}
    \vspace{0.0cm}
   \includegraphics[width=60mm]{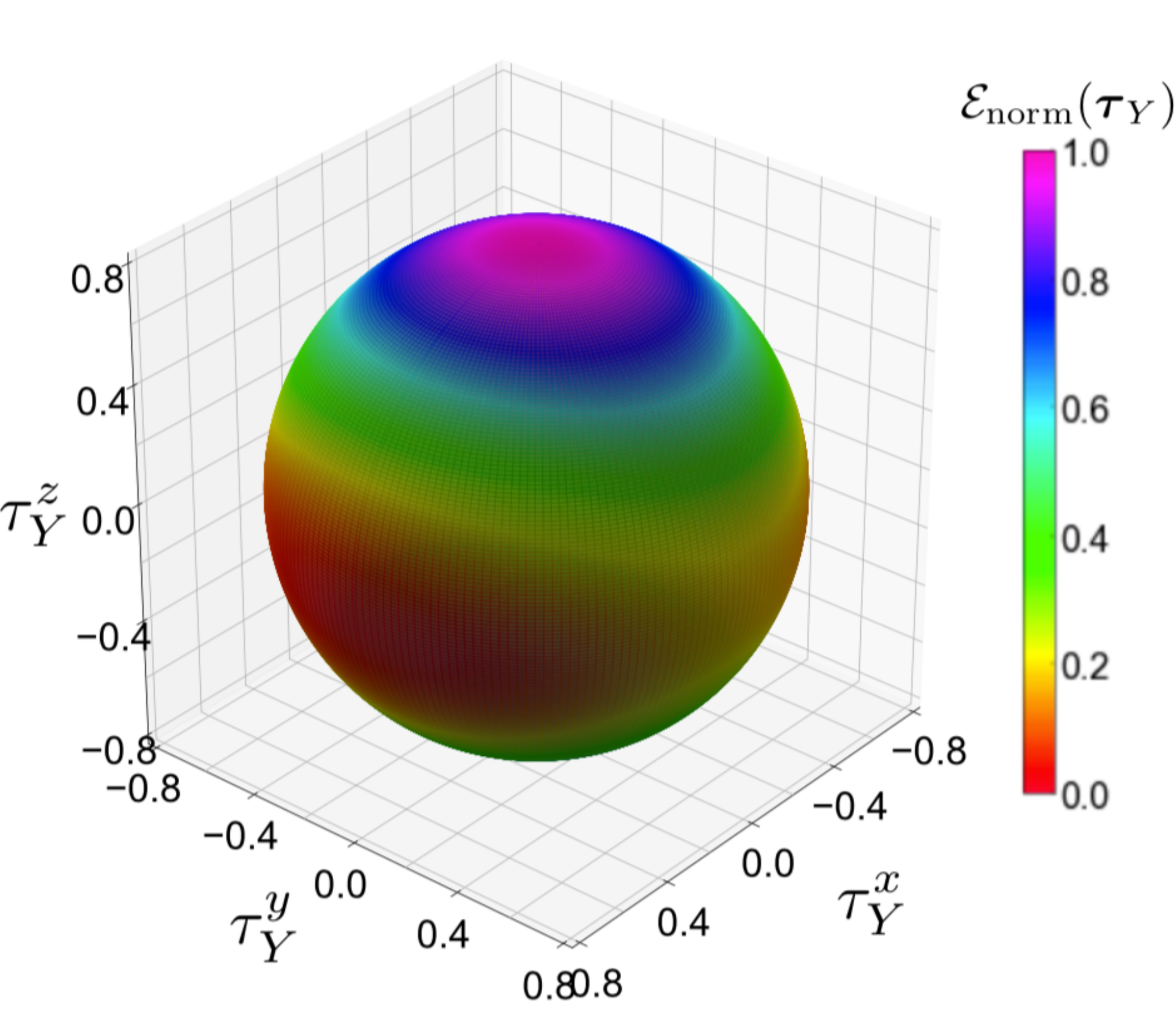} 
  \caption{Total energy $\mathcal{E}({\btau_Y})$ (Eq.~\eqref{eq:H_eff_energy}) as a function of angle of ${\btau_Y}$ on the sphere with radius $r$.
  The color represents the normalized energy, $\mathcal{E}_{\rm norm}({\btau_Y})$.
  This corresponds to the quench from $\Delta_{\rm cf}=2.0$ and $T=0.3$ to $\Delta_{\rm cf}=1.5$. Here, we use  $v=1$, $v'=0$, $U=6$, $J=1$, and $I=\frac{ \Delta_{\rm cf}}{1.5} $ and apply $h_{\rm seed}=\sqrt{2}\cdot 10^{-3}$ to $\sum_i  \htau_{Y,i}^x$. The parameters of the effective model after the quench become $h_z=-0.0479$, $z_nJ_x=0.281$, $z_nJ_y=0.192$, $z_nJ_z=0.0396$, $m=0.0483$, and $r=0.807$.  }
  \label{fig:Energy_voff_zero}
\end{figure}
%%%%%%%%%%%%%%%%%%%%%%%%%%%%%%%%%%%%%%%%%%

We note that within MF theory, the density matrix $\hat{\mathcal{P}}$ of the system (not the single-particle density matrix) can be expressed 
as the direct product of the density matrices at each site $\hat{\mathcal{P}}_i$ as $\hat{\mathcal{P}} = \otimes_i \hat{\mathcal{P}}_i$. 
Since the time evolution of $\hat{\mathcal{P}}_i$ is determined by the local Hamiltonian $\hH^{\rm MF}_i$, the local entropy defined as $S_i \equiv - \hat{\mathcal{P}}_i \ln \hat{\mathcal{P}}_i$ is independent of time.
In this sense, the MF dynamics conserves the local entropy. 
For example, if the initial state is a pure low-spin state, $S_i$ remains zero and $\hat{\mathcal{P}}_i(t)$ represents a pure state at any time in the evolution.
Hence, the MF dynamics after a quench from a low-spin to a high-spin state has a built-in cooling related to the conservation of the entropy.
In the pseudo-spin picture described by Eq.~\eqref{eq:H_eff_simple}, this conservation is connected to the conservation of the norm of  $\btau_Y$.
Because of the conservation of the norm, if   $\btau_Y$ can deviate from the $(0,0,-1)$ direction, i.e. the low-spin state, the $x$ and $y$ components can become large.

\subsection{D. 2. Model with $v=1,\;\; v'=0$}

In Fig.~\ref{fig:phase_voff_zero}(a), we show the MF phase diagram of the effective model with a small seed field coupled  to $\sum_{i}\htau_{Yi}^x$ in the region of the high-spin/low-spin crossover.
There emerges an excitonic order near the high-spin/low-spin crossover, where only ${\rm Re}\langle \hat{d}^\dagger_Y \hh\rangle$ is nonzero.
${\rm Re}\langle \hat{d}^\dagger_Y \hh\rangle$, or in other words $\htau_{\Gamma}^x$, is favored since in this case we have $J_x >  |J_y|$ in the language of the effective Hamiltonian \eqref{eq:H_eff2}.
The probability of high-spin states ($n_d$) is shown in Fig.~\ref{fig:phase_voff_zero}(b), and the results are qualitatively the same as in Fig.~\ref{fig:phase_v_voff}.

Now, we consider a quench from a low-spin state at a high temperature. 
Figure~\ref{fig:Quench_voff_zero} shows the time evolution of the expectation value of  the pseudo-spins and the $S=1$ spins after the quench from $\Delta_{\rm cf}=2.0$ and $T=0.3$ to $\Delta_{\rm cf} = 1.5$ with the seed field $h_{\rm seed} = \sqrt{2}\cdot 10^{-3}$.
In stark contrast to the case of $v'=1$, there is no enhancement of the excitonic order and the system remains in the low-spin state, which is consistent with the DMFT results.

This can be understood in terms of  the simplified effective model \eqref{eq:H_eff_simple} and its $\btau_Y$-dependent energy profile \eqref{eq:H_eff_energy}.
With $v'=0$, it turns out that $J_x$ and $J_y$ are comparable and both positive. 
Thus, after the quench, there emerges a 
ring shaped low energy region around $\tau_Y^z\simeq 0$, as exemplified in Fig.~\ref{fig:Energy_voff_zero}.
Thus the enhancement of the excitonic order is energetically not allowed,
although the entropy of the system is low.

%\begin{thebibliography}{99}
%\bibitem{Kunes_2014} J. Kunes and P. Augustinsky, Phys. Rev. B {\bf 89},115134 (2014). 
%\bibitem{Kunes_2015} J. Kunes, J. Phys.: Condens. Matter {\bf 27}, 333201 (2015).
%\bibitem{Nasu_2016} J. Nasu, T. Watanabe, M. Naka, and S. Ishihara, Phys. Rev. B {\bf 93}, 205136 (2016).
%\end{thebibliography}


\begin{thebibliography}{99}
\bibitem{Fausti_2011} D. Fausti, R. I. Tobey, N. Dean, S. Kaiser, A. Dienst, M. C. Hoffmann, S. Pyon, T. Takayama, H. Takagi, and A. Cavalleri, Science {\bf 331}, 189 (2011).
\bibitem{Kaiser_2014} S.Kaiser, C. R. Hunt, D. Nicoletti, W. Hu, I. Gierz, H. Y. Liu, M. Le Tacon, T. Loew, D. Haug, B. Keimer, and A. Cavalleri, Phys. Rev. B {\bf 89}, 184516 (2014).
\bibitem{Mitrano_2016} M. Mitrano, A. Cantaluppi, D. Nicoletti, S. Kaiser, A. Perucchi, S. Lupi, P. Di Pietro, D. Pontiroli, M. Ricco, S. R. Clark, D. Jaksch, and A. Cavalleri, Nature {\bf 530}, 461 (2016).
\bibitem{Denny_2015} S. J. Denny, S. R. Clark, Y. Laplace, A. Cavalleri, and D. Jaksch, Phys. Rev. Lett {\bf 114}, 137001 (2015).
\bibitem{Sentef2016} M. A. Sentef, A. F. Kemper, A. Georges, and C. Kollath, Phys. Rev. B {\bf 93}, 144506 (2016).
\bibitem{Okamoto_2016} J.-i. Okamoto, A. Cavalleri, and L. Mathey, Phys. Rev. Lett. {\bf 117}, 227001 (2016).
\bibitem{Kennes2017} D. M. Kennes, E. Y. Wilner, D. R. Reichman, and A. J. Millis, Nature Physics {\bf 13}, 479 (2017).
\bibitem{Mazza_2017}  G. Mazza and A. Georges, Phys. Rev. B 96, 064515 (2017).
\bibitem{Werner2018} P. Werner, H. Strand, S. Hoshino, Y. Murakami, and M. Eckstein, Phys. Rev. B {\bf 97}, 165119 (2018).
\bibitem{Nava_2018} A. Nava, C. Giannetti, A. Georges, E. Tosatti, and M. Fabrizio, Nature Physics {\bf 14}, 154 (2018).
\bibitem{Kaneko_2019}  T. Kaneko, T. Shirakawa, S. Sorella, and S. Yunoki, Phys. Rev. Lett. {\bf 122}, 077002 (2019).
\bibitem{Werner_2019b}  P. Werner, J. Li, D. Golez, and M. Eckstein, Phys. Rev. B {\bf 100}, 155130 (2019).
\bibitem{Kennes_2019} D. M. Kennes, M. Claassen, M. A. Sentef, and C. Karrasch, Phys. Rev. B {\bf 100}, 075115 (2019).
\bibitem{Tindall_2019}  J. Tindall, B. Buca, J. R. Coulthard, and D. Jaksch, Phys. Rev. Lett. 123, 030603 (2019).
\bibitem{Buzzi_2020} M. Buzzi, D. Nicoletti, M. Fechner, N. Tancogne-Dejean, M. A. Sentef, A. Georges, M. Dressel, A. Henderson, T. Siegrist, J. A. Schlueter, K. Miyagawa, K. Kanoda, M. -S. Nam, A. Ardavan, J. Coulthard, J. Tindall, F. Schlawin, D. Jaksch, and A. Cavalleri, arxiv:2001.05389 (2020).
\bibitem{Mor_2017} S. Mor, M. Herzog, D. Golez, P. Werner, M. Eckstein, N. Katayama, M. Nohara, H. Takagi, T. Mizokawa, C. Monney, and J. St\"ahler, Phys. Rev. Lett. {\bf 119}, 086401 (2017).
\bibitem{Murakami_2017} Y. Murakami, D. Golez, M. Eckstein, and P. Werner, Phys. Rev. Lett. {\bf 119}, 247601 (2017).
\bibitem{Ohta_2018} T. Tanabe, K. Sugimoto, and Y. Ohta, Phys. Rev. B {\bf 98}, 235127 (2018).
\bibitem{Yonemitsu_2018} Y. Tanaka, M. Daira, and K. Yonemitsu, Phys. Rev. B {\bf 97}, 115105 (2018).
\bibitem{Stefanucci_2019} E. Perfetto, D. Sangalli, A. Marini, and G. Stefanucci,  Phys. Rev. Materials {\bf 3}, 124601 (2019).
\bibitem{Murotani_2019} Y. Murotani, C. Kim, H. Akiyama, L. N. Pfeiffer, K. W. West, and R. Shimano,  Phys. Rev. Lett. {\bf 123}, 197401 (2019).
\bibitem{Murakami_2020} Y. Murakami, M. Sch\"uler, S. Takayoshi, and P. Werner, Phys. Rev. B {\bf 101}, 035203 (2020).
\bibitem{Ono_2017} A. Ono and S. Ishihara, Phys. Rev. Lett. {\bf 119}, 207202 (2017).
\bibitem{Werner_2019a} P. Werner, M. Eckstein, M. M\"uller, and G. Refael, Nature Comm. {\bf 10}, 5556 (2019).
\bibitem{Bernier_2009} J.-S. Bernier, C. Kollath, A. Georges, L. De Leo, F. Gerbier, C. Salomon, C. K\"ohl, Phys. Rev. A {\bf 79}, 061601(R) (2009).
\bibitem{Fabrizio2018} M. Fabrizio, Phys. Rev. Lett. {\bf 120}, 220601 (2018).
\bibitem{Mazurenko_2017} A. Mazurenko, C. S. Chiu, G. Ji, M. F. Parsons, M. Kanasz-Nagy, R. Schmidt, F. Grusdt, E. Demler, D. Greif and M. Greiner, Nature {\bf 545}, 462 (2017).
\bibitem{Chiu_2018} C. Chiu, G. Ji, A. Mazurenko, D. Greif, and M. Greiner, Phys. Rev. Lett. {\bf 120}, 243201 (2018).
\bibitem{Kaneko_2012} T. Kaneko, K. Seki, and Y. Ohta, Phys. Rev. B {\bf 85}, 165135 (2012).
\bibitem{Kunes_2014} J. Kunes and P. Augustinsky, Phys. Rev. B {\bf 89},115134 (2014). 
\bibitem{Kunes_2015} J. Kunes, J. Phys.: Condens. Matter {\bf 27}, 333201 (2015).
\bibitem{Hoshino_2016} S. Hoshino and P. Werner, Phys. Rev. B {\bf 93}, 155161 (2016).
\bibitem{Nasu_2016} J. Nasu, T. Watanabe, M. Naka, and S. Ishihara, Phys. Rev. B {\bf 93}, 205136 (2016).
\bibitem{Georges_1996} A. Georges, G. Kotliar, W. Krauth, and M. J. Rozenberg, Rev. Mod. Phys. {\bf 68}, 13 (1996).
\bibitem{Aoki_2014} H. Aoki, N. Tsuji, M. Eckstein, M. Kollar, T. Oka, and P. Werner,
Rev. Mod. Phys. {\bf 86}, 779 (2014).
\bibitem{Keiter1971} H. Keiter and J. Kimball, Int. J. Magnetism {\bf 1}, 233 (1971).
\bibitem{Eckstein2010} M. Eckstein and P. Werner, Phys. Rev. B {\bf 82}, 115115 (2010).
\bibitem{Werner_2007} P. Werner and A. J. Millis, Phys. Rev. Lett. {\bf 99}, 126405 (2007).
\bibitem{footnote_op}{This OP is related to the spin-orbital OP $o^{\mu\lambda}=\sum_{\gamma\gamma'\sigma\sigma'} \sigma^\mu_{\gamma\gamma'}\sigma^\lambda_{\sigma\sigma'}\langle c^\dagger_{\gamma\sigma}c_{\gamma'\sigma'}\rangle$ \cite{Hoshino_2016}, for example $\phi^Y = \tfrac i2 o^{yy}$.} 
\bibitem{footnote_phiz} If we consider the EI with dominant $\phi^Z$ the situation is opposite.
\bibitem{Eckstein2009} M. Eckstein, M. Kollar, and P. Werner, Phys. Rev. Lett. {\bf 103}, 056403 (2009).
\bibitem{Schiro2010} M. Schiro and M. Fabrizio, Phys. Rev. Lett. {\bf 105}, 076401 (2010).
\bibitem{Mardegan_2020} J. R. L. Mardegan {\it et al.}, arXiv:2002.12214 (2020).
\bibitem{Nessi} M. Sch\"uler, D. Gole\v{z}, Y. Murakami, N. Bittner, A. Hermann, H. U. Strand, P. Werner, and M. Eckstein, arXiv:1911.01211 (2019).
\bibitem{comment} One can see that there exists an energetically allowed point at $\btau_Y = (0,C,D)$ for $E_\text{tot}(0,0,|\tau_Y|)>E_\text{tot}(0,0,-|\tau_Y|)$. 
 If $E_\text{tot}(0,0,|\tau_Y|)<E_\text{tot}(0,0,-|\tau_Y|)$, this point is no longer energetically allowed, while one exists at $\btau_Y = (C,0,D)$. 
\end{thebibliography}
\end{document}